%% file: paper.tex
\documentclass[aps,prb,reprint,amsmath,amssymb,floatfix,superscriptaddress,showkeys]{revtex4-1}
\usepackage{graphicx}
\usepackage{color}
\usepackage{soul}
\usepackage{amssymb, amsmath}
\usepackage{threeparttable}

\begin{document}
\title[Metallicities of GCs]{Determining Metallicities of Globular Clusters using Simulated Integrated Spectra and Bayesian Statistics}
\author{C. Euler}
\email{eulerch@uni-mainz.de.}
\affiliation{Institut f\"ur Physik, Johannes Gutenberg-Universit\"at Mainz, Staudinger Weg 7, 55128 Mainz, Germany}

\date{\today}
\label{firstpage}
\begin{abstract}
	Using Monte Carlo simulations of globular clusters we developed a method separating metallicity effects from age effects on observed integrated ugriz colors. We demonstrate that these colors do not evolve with time significantly after an age of 4~Gyr and use Bayesian statistics to calculate a probability distribution function of the metallicity. We tested the method using the M31 globular cluster system and then applied to explain the observed color bimodality in globular cluster sets and tidal effects on it. We show that the color bimodality is an effect of a nonlinearity in the color-metallicity relation caused by stellar dynamics on the Giant Branch, that colors including only the UV show a weaker bimodality than those subtracting from visual bands and that cluster sets with a distinct bimodality are in principle older than those with only a weak bimodal distribution. Furthermore a bimodal color distribution of coeval clusters implies a bimodal metallicity distribution, but a unimodal color distribution does not imply a unimodal metallicity distribution. The tidal field can finally shift the modes of the color distribution and therefore cause a bimodal color distribution. This work presents results obtained between 2011 and 2012 in the Astronomisches Rechen-Institut, Zentrum f\"{u}r Astronomie der Universit\"{a}t Heidelberg, M\"{o}nchhofstra{\ss}e 12-14, 69120 Heidelberg, Germany.
\end{abstract}
%

\maketitle

\section{Introduction}
\label{sec:intro}
Globular clusters are commonly believed to consist of one distinct population of stars. Groups of globular clusters seem to evolve differently in a similar environment, producing at least two so called branches of clusters with different colors (see e.g. \citet{Zinn85} and \citet{Peng11}). To a smaller extent the color-bimodality could be influenced by the tidal environment: The radial distribution of globular clusters around giant elliptical galaxies like M87 shows that red, and presumably metal rich, clusters are more concentrated, while blue clusters are spread out further (\citet{Harris10a}, \citet{Peng11}). The color-bimodality is thought to be caused by the metallicity distribution of the clusters, which is commonly found either by spectroscopy or from integrated light by fitting of SSP models of clusters to observational data. Here, the transformation between color and metallicity is strongly nonlinear, the nonlinearity possibly being caused by stellar evolution. Because a cluster's color both develops with time and depends on the initial metallicity, a degeneracy appears prohibiting a measurement of the metallicity distribution from the integrated color-magnitude parameter space. Still it is usually assumed -- and the results of this paper indicate that this view is at least in principle correct -- that metal-rich clusters are redder than their metal-poor counterparts.

This paper describes a method to explain these observations and to resolve the age-metallicity degeneracy. To achieve this goal we have produced a integrated color-metallicity relation for a wide range of metallicities from which the metallicity of an observed cluster can be inferred. This catalog also sheds light on the origins of the color-bimodality. This work first explains the methods used to obtain the catalog. In section \ref{sec:evolution} we show how ugriz colors develop with time and demonstrate that after an age of 4~Gyr there is only very little development of color with time. Based on this in the fourth and fifth section a method to separate metallicity from age is developed and further tested using M31 data. Section \ref{sec:vir} describes the application to the Virgo galaxy cluster globular cluster population.

\section{Numerical models}
\label{sec:models} 

The simulation method is based on a Monte Carlo technique first implemented by \citet{Henon72a} and \citet{Henon72b}, then later refined by \citet{Stodolkiewicz82} and more recently by \citet{Giersz98}. It is based on the two-body relaxation limit of the Fokker-Planck approximation of the collisional Boltzmann equation and therefore only considers statistical properties of a cluster, but is very well adjusted to produce physical results. Despite the statistical nature of the simulation it is possible to acquire physical information about single stars, specifically temperature, radius, metallicity and luminosity, which are needed to compute spectra. In order to do so it assumes a sufficiently large ($>10,000$ particles) spherically symmetric system older than the relaxation time. Compared to other simulation techniques like direct N-Body codes it is very fast as it scales linearly or at most quadratic with $N$, so that large numbers of simulations can be run to produce results of statistical significance. Stellar evolution is included following the formalism of \citet{Hurley00} and \citet{Hurley02}.

As a Monte Carlo simulation is statistical, all runs with identical random seed should produce the same output. For error estimation, 16 simulations with identical initial conditions but different random seeds chosen randomly have been performed. The results for the SDSS ugriz bands are displayed in Tab. \ref{tab:error}. The errors in luminosity are below one per cent throughout all times and wavelength bands. For colors, error propagation though leads to noticeable uncertainties, e.g. for $g-i$ color the error ranges between $\Delta (g-i)\approx 0.01$ at $t=1~\mathrm{~Gyr}$ and $\Delta (g-i)\approx 0.04$ at $t=12~\mathrm{~Gyr}$. As observational errors will be at least in this range or even larger, the predictive power of the simulation data in terms of color is limited. This fact will prove to be useful in the analysis of the effect of a cluster's dynamic environment in the next section.

\begin{table*}
\centering
\caption{Absolute luminosity in the ugriz bands with corresponding standard deviations for 1, 6, 10 and 12~Gyr cluster age.}
\begin{minipage}{126mm}
\begin{tabular}[c]{ccccccccccc}
  \hline
  \multicolumn{11}{c}{Statistical errors of simulated ugriz colors}\\
  \hline
   Time &  $L_u$ &  $\sigma_u$ & $L_g$ &  $\sigma_g$ &  $L_r$ &  $\sigma_r$ & $L_i$ &  $\sigma_i$ & $L_z$ &  $\sigma_z$\\
  \hline
  1~Gyr & -6.407 & 0.005 & -7.757 & 0.006 & -8.172 & 0.009 & -8.322 & 0.013 & -8.533 & 0.028\\\par
  6~Gyr & -3.871 & 0.007 & -5.578 & 0.011 & -6.392 & 0.016 & -6.702 & 0.022 & -7.010 & 0.038\\\par
  10~Gyr & -3.075 & 0.019 & -4.942 & 0.018 & -5.827 & 0.025 & -6.158 & 0.033 & -6.467 & 0.051\\\par
  12~Gyr & -2.680 & 0.015 & -4.616 & 0.022 & -5.502 & 0.027 & -5.837 & 0.028 & -6.167 & 0.047\\\par
  \hline
\end{tabular}
\end{minipage}
\label{tab:error}
\end{table*}

All luminosity and color data produced is based on the simulation of integrated spectra using Galev (\textbf{Gal}axy \textbf{Ev}olution) (\citet{Kotulla09}). A star is modeled using a one-zone model without any spatial resolution. While stellar population synthesis or differential synthesis algorithms are only able to find a best fit for observational data from linear combination of spectra, Galev has the advantage of producing results for arbitrary star formation histories. For a Simple Stellar Population (SSP) which the globular clusters are in this context assumed to be the star formation history is a delta distribution. \citet{Schulz02} and \citet{Anders03} were able to reproduce globular clusters as SSPs well in an early stage of Galev. They showed that a color-age relation is only valid at one metallicity and a color-metallicity relation is only valid at a distinct age (\citet{Schulz02}). Galev uses isochrones by the Padova group (e.g. \citet{Girardi02}) for metallicities of [Fe/H] in the range of $-1.7$, $-0.7$, $-0.4$, $0.0$ and $0.4$ and also include thermally pulsing AGB stars (\citet{Schulz02}). For emission line contributions a set of zero age main sequences (ZAMS) is included up to masses of $120 \mathrm{M}_\odot$.

A library of stellar spectra is required to compute a spectrum from a given effective temperature, surface gravity and metallicity. For normalization in the process of combining spectra of SSPs the luminosity of a star is needed. Temperature and radius (and following from it, surface gravity) are used to distinguish between stellar types of similar temperature but different evolutionary stage. These spectra are then convolved with filter transmission curves used in observations (e.g. the SDSS ugriz filter set) and normalized to any magnitude system. In this case, the AB system is used.

Galev uses the BaSeL library of model atmospheres published by \citet{Lejeune97}, \citet{Lejeune98a} and \citet{Lejeune98b} based on \citet{Kurucz70}, \citet{Kurucz93} and others. The wavelength spans 90~\r{A} to $160~\mu\mathrm{m}$ from the extreme ultraviolet to the far infrared with a resolution of 20~\r{A} in the ultraviolet and optical and 50~\r{A}$\dots$100~\r{A} in the infrared. Galev uses a large set of elements to compute spectra with data taken from \citet{Woosley95} for high-mass stars and from \citet{Hoek97} for low-mass stars.

For 100 in logarithm equidistant metallicities between $0.02$ (solar) and $0.0002$ simulations with compared to the relaxation time high temporal resolution of $50\,\mathrm{Myr}$ up to 12~Gyr of age have been run. A wide range of the parameter space has been explored, but, as long as not stated otherwise, results described have been obtained using models with $N_0=50,000$, an initial King profile parameter of $W_0=6$, an initial binary fraction of 10\% and an initial tidal radius of 160 pc.

\section{Time Evolution of ugriz Colors}
\label{sec:evolution}

\begin{figure*}
  \centering
  \includegraphics[width=0.45\textwidth]{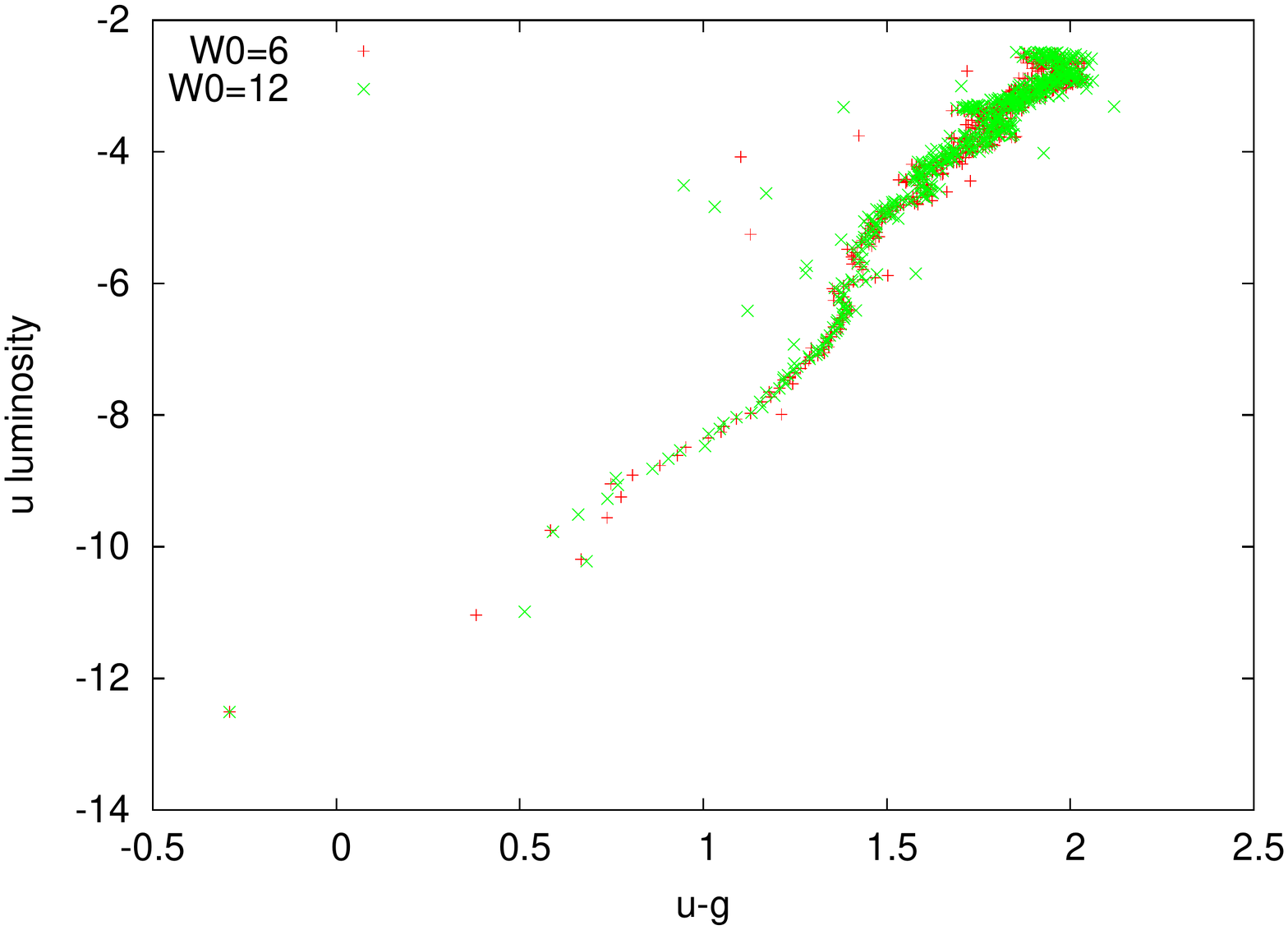}
  \includegraphics[width=0.45\textwidth]{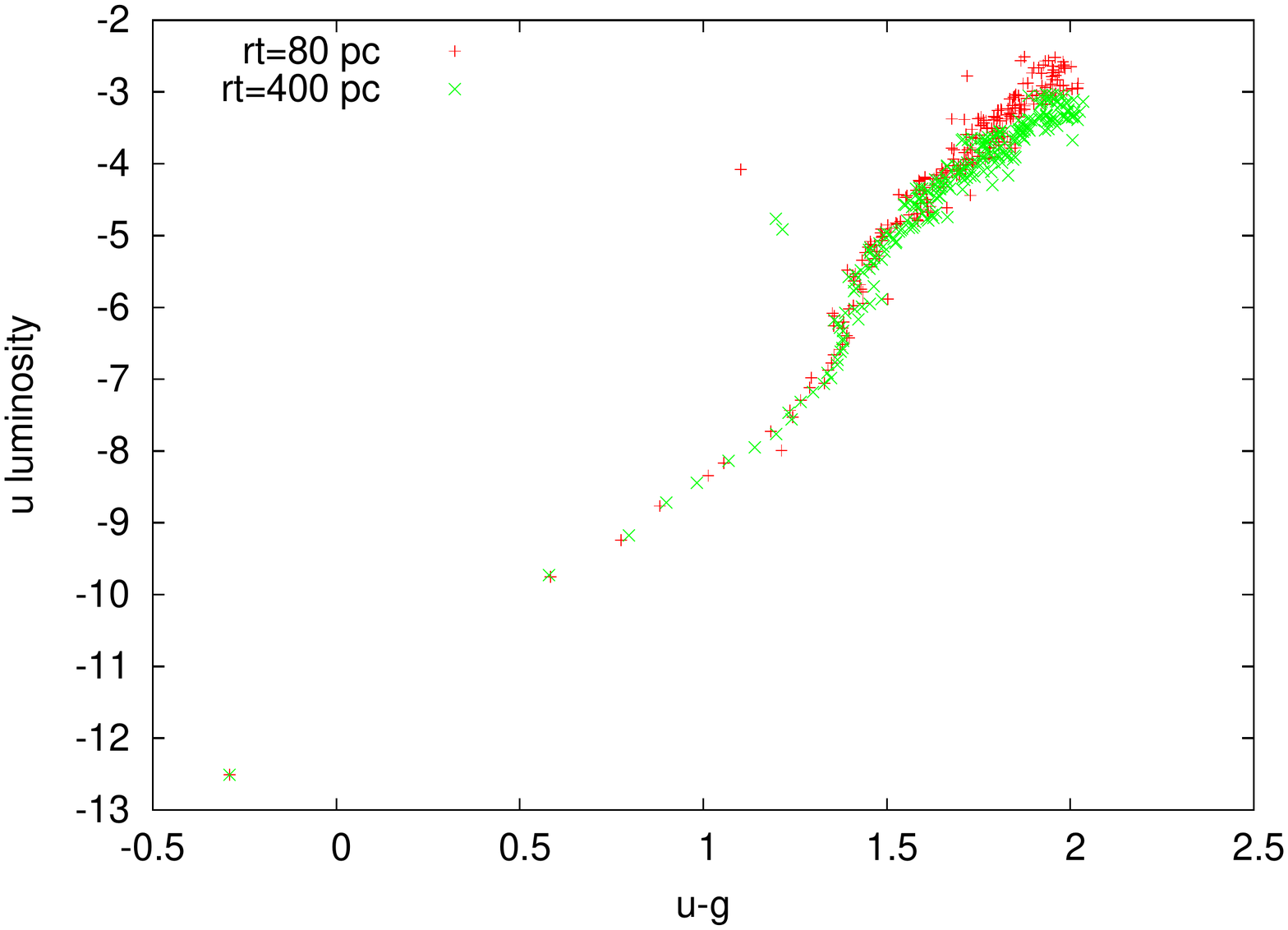}\\
  \includegraphics[width=0.45\textwidth]{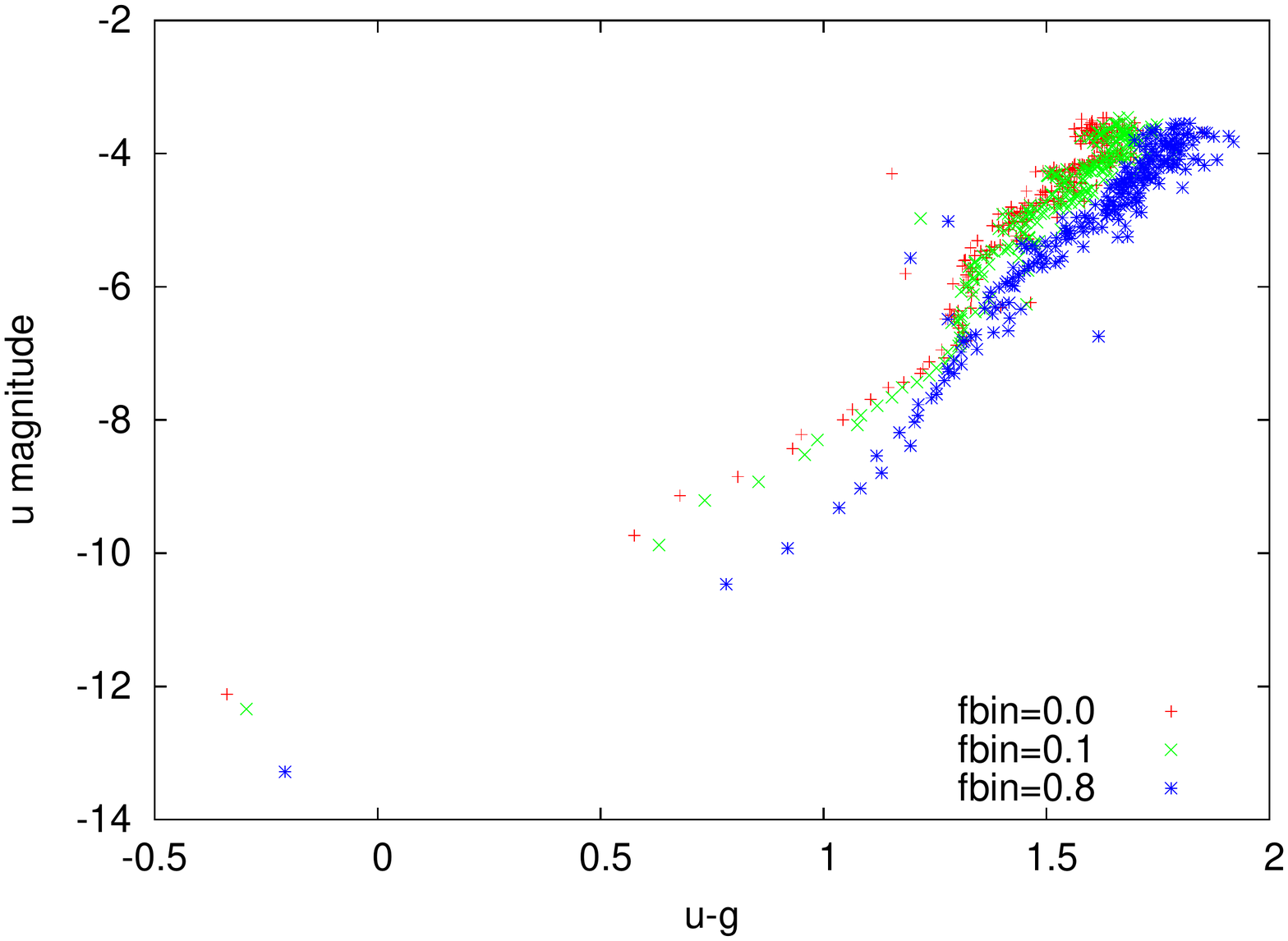}
  \includegraphics[width=0.45\textwidth]{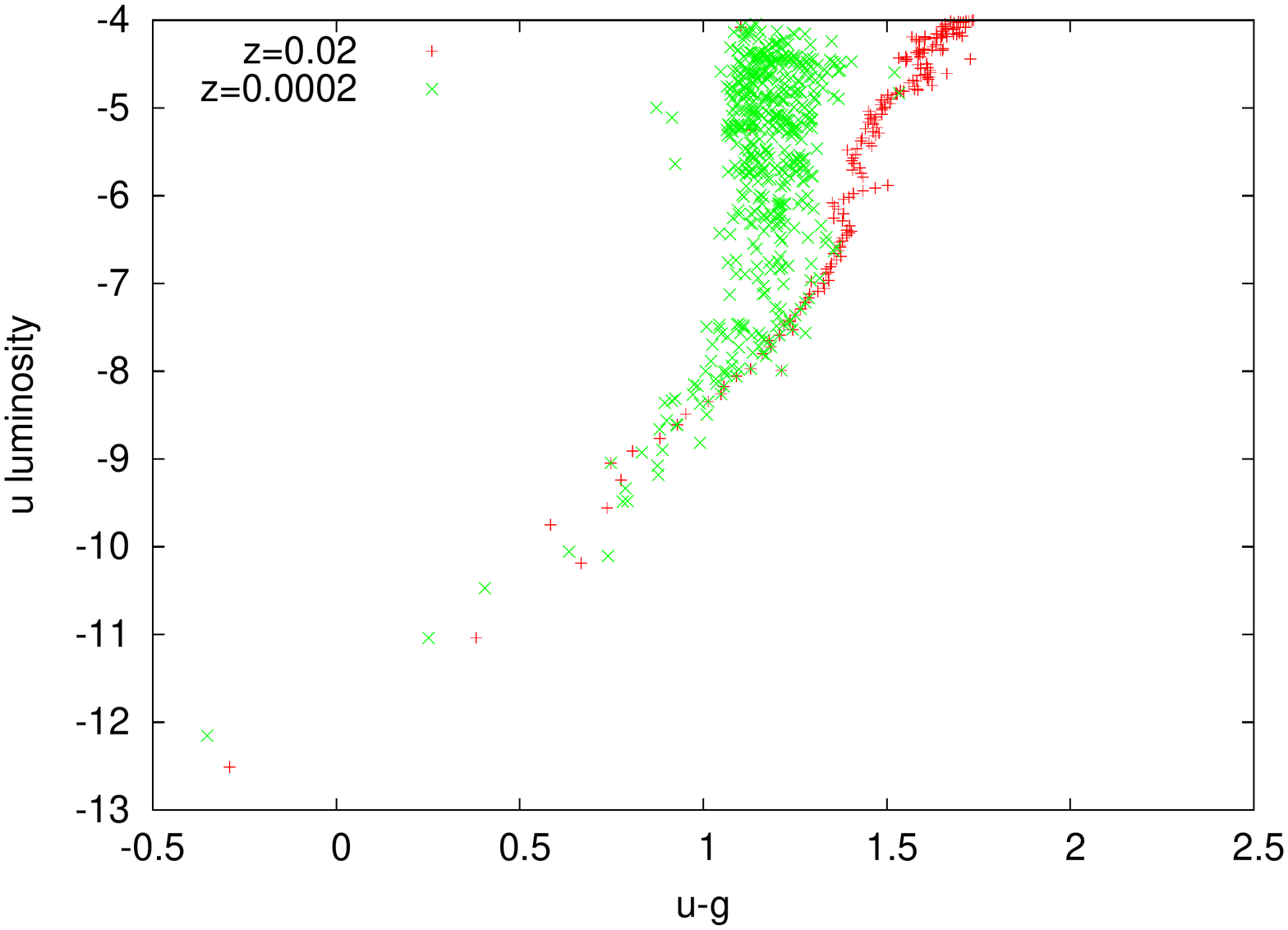}
  \caption{Exemplarily the trajectory in  u absolute luminosity vs. u-g color space is shown for simulations of $N=50,000$ particles with different initial King profile parameter (top left), tidal radius (top right), binary fraction (bottom left) and metallicity (bottom right).}
  \label{fig:highres}
\end{figure*}

In this section the sensitivity of a cluster's trajectory in integrated color-magnitude space for change of $W_0$, $r_\mathrm{tid}$, $f_\mathrm{bin}$ and $z_\mathrm{ini}$ is studied.
The effect of the initial concentration should be very low, because relaxation processes should very quickly erase all memory of the initial state, as they elapse on the relaxation time scale. Any differences between clusters with different initial concentrations should have vanished after some 100 Myr. The top left panel of Fig. \ref{fig:highres} shows the trajectory of a cluster's integrated properties in color magnitude space for different choices of $W_0$ (as above the extreme value of $W_0=12$ is adopted as well as a Plummer-like model with $W_0=6$). The behavior of both curves is very similar and the choice of the initial concentration is, if at all, only important before $t=500\mathrm{Myr}$ at $u-g\approx 1.2$. After 1~Gyr ($g-i\approx 1.5$) the intrinsic scatter increases and outliers to the left appear. These very blue and slightly brighter cluster states seem to be caused by UV bright stars. As UV bright stars have \lq life times\rq\, of $10^4$ to several $10^5$ years, it is a coincidence that they appeared in the simulation output, because the time step chosen for simulations is 20 Myr. These stars disrupt the smooth trajectory along which the cluster evolves in the chosen CMD. The large scatter in color at later times is statistical (see error values in table \ref{tab:error}) and this leads to the conclusion that the initial concentration is not relevant for color evolution after several $100\,\mathrm{Myr}$.

Changing the binary fraction will have an effect, as the initial number of bodies in the system counts a binary as a single body, which consequently is brighter than a single star. Therefore, increasing the binary fraction leads to higher luminosity. After initial relaxation processes have ended and soft binaries are disrupted, the binary fraction stays constant and only a systematic color difference should remain. This color difference (clusters with more binaries are slightly redder) is caused by binaries tending to evolve faster as they exchange mass, if their separation is small enough for their Roche lobes to connect, and therefore shift the total cluster color towards the red side of the diagram. A different choice of the initial binary fraction as shown exemplarily in the bottom left pannel of Fig. \ref{fig:highres} has this obvious effect, even though with decreasing error in luminosity of the simulation with time the deviations of different runs disappear.

\begin{figure}
  \centering
  \includegraphics[width=0.5\textwidth]{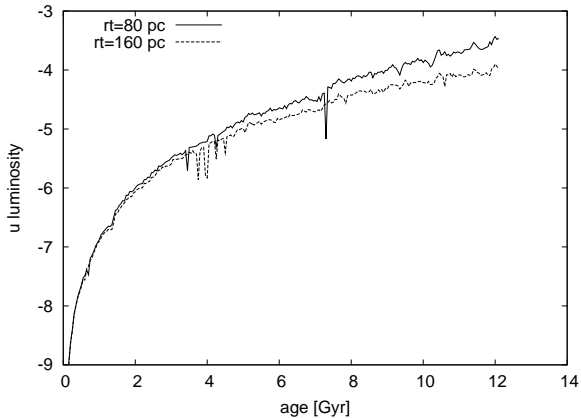}
  \caption{The effect of the tidal radius on, exemplarily, u magnitude. Two clusters with the usual $N_0=50,000$, $z_\mathrm{ini}=0.02$ and $f_\mathrm{bin}=0.1$ but tidal radii of 80~pc and 160~pc are compared in their development of u luminosity.}
  \label{fig:rtid-lum}
\end{figure}

The choice of initial metallicity should very much influence the color of the cluster, because stellar evolution depends on metallicity and high-metallicity stars evolve faster than their low-metallicity counterparts. Therefore, a strong difference between clusters with different metallicities is to be expected. The bottom right panel of Fig. \ref{fig:highres} displays cluster trajectories for two different choices of initial metallicity. The intrinsic spread increases dramatically with decreasing metallicity and also, but less, with time. Red high-z clusters quickly move towards a region of $u-g > 1.25$ and then develop more in luminosity than in color. With decreasing metallicity this effect diminishes and the spread in color grows. Still, a multi-modality remains: Young sets of clusters have a spread dominated by low-z clusters and have fairly monotonous high-z contributions. After approximately one Billion years the multi-modality begins developing its shape and increases in distinctiveness with time.

Changing the tidal radius (see top right panel of Fig. \ref{fig:highres}) has very little effect for the first Billion years. As expected, there are differences at later stages of cluster evolution, as a small tidal radius leads to a higher escape rate and therefore to less luminosity and through increased mass segregation also to bluer clusters. This effect is insignificant in color in terms of the accuracy of the simulation. In luminosity though it is very relevant and can significantly influence the integrated light and therefore measurements of distance and related quantities (see Fig. \ref{fig:rtid-lum}). Two clusters with the usual $N_0=50,000$, $z_\mathrm{ini}=0.02$ and $f_\mathrm{bin}=0.1$ but tidal radii of 80 pc and 160 pc are compared exemplarily in their development of u luminosity. Any quantity derived from luminosity will be underestimated due to the difference in these curves. To correct for this effect a method to measure the initial tidal radius is needed, so that the systematic error can be accounted for.

\section{Disentangling Age and Metallicity: GloMAGE}
\label{sec:glomage}
\begin{figure*}
  \includegraphics[width=0.45\textwidth]{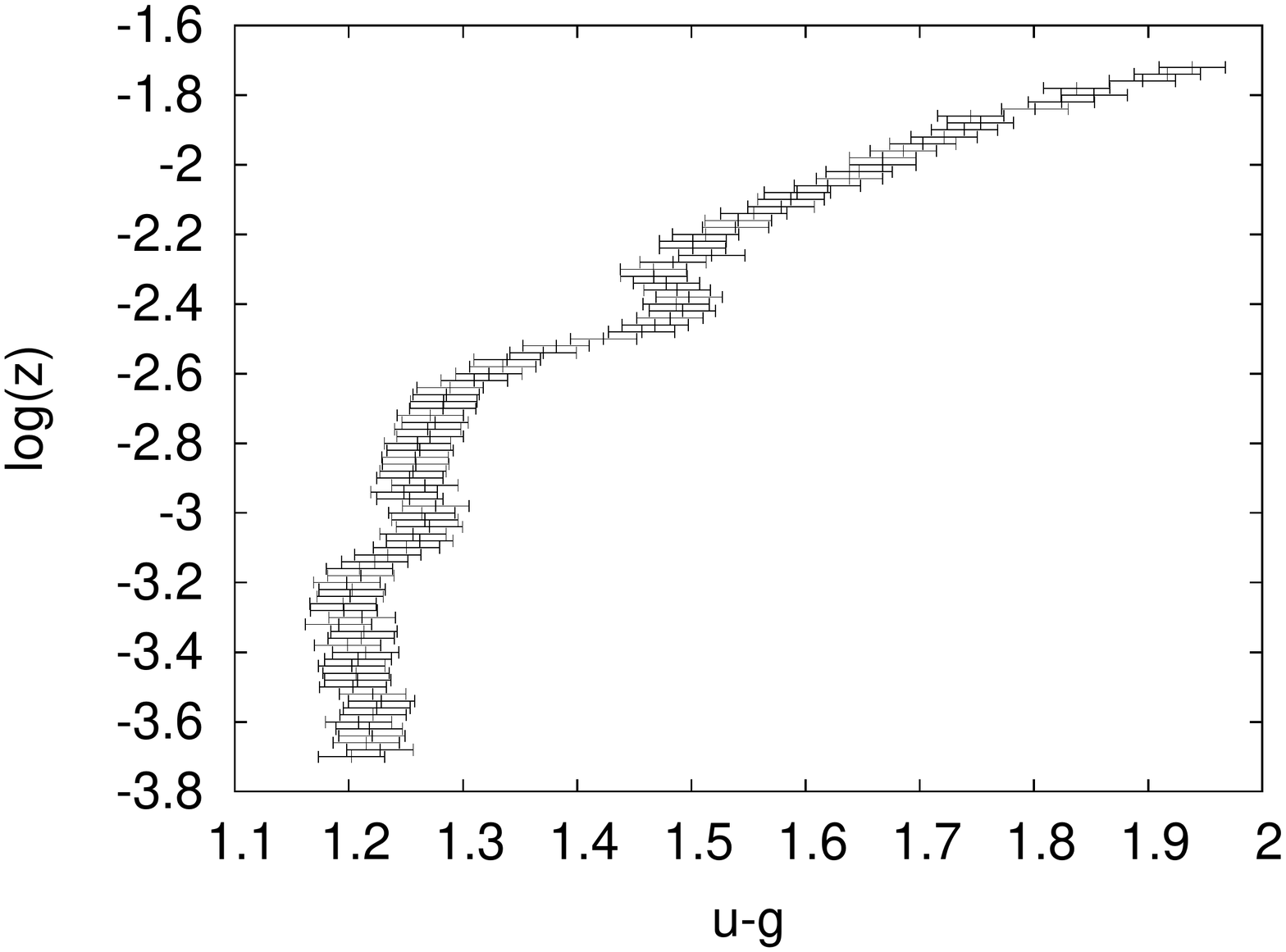}
  \includegraphics[width=0.45\textwidth]{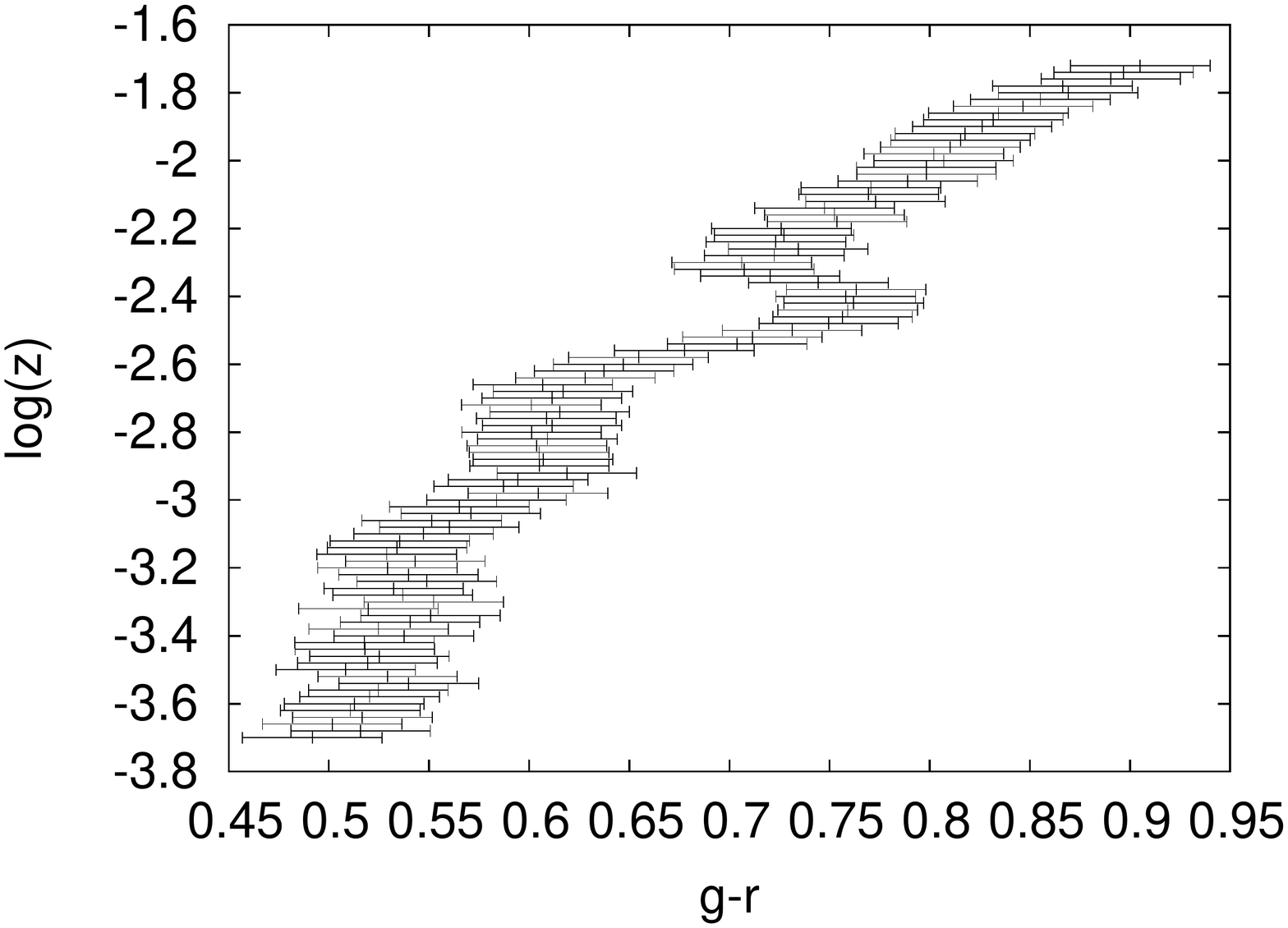}\\
  \includegraphics[width=0.45\textwidth]{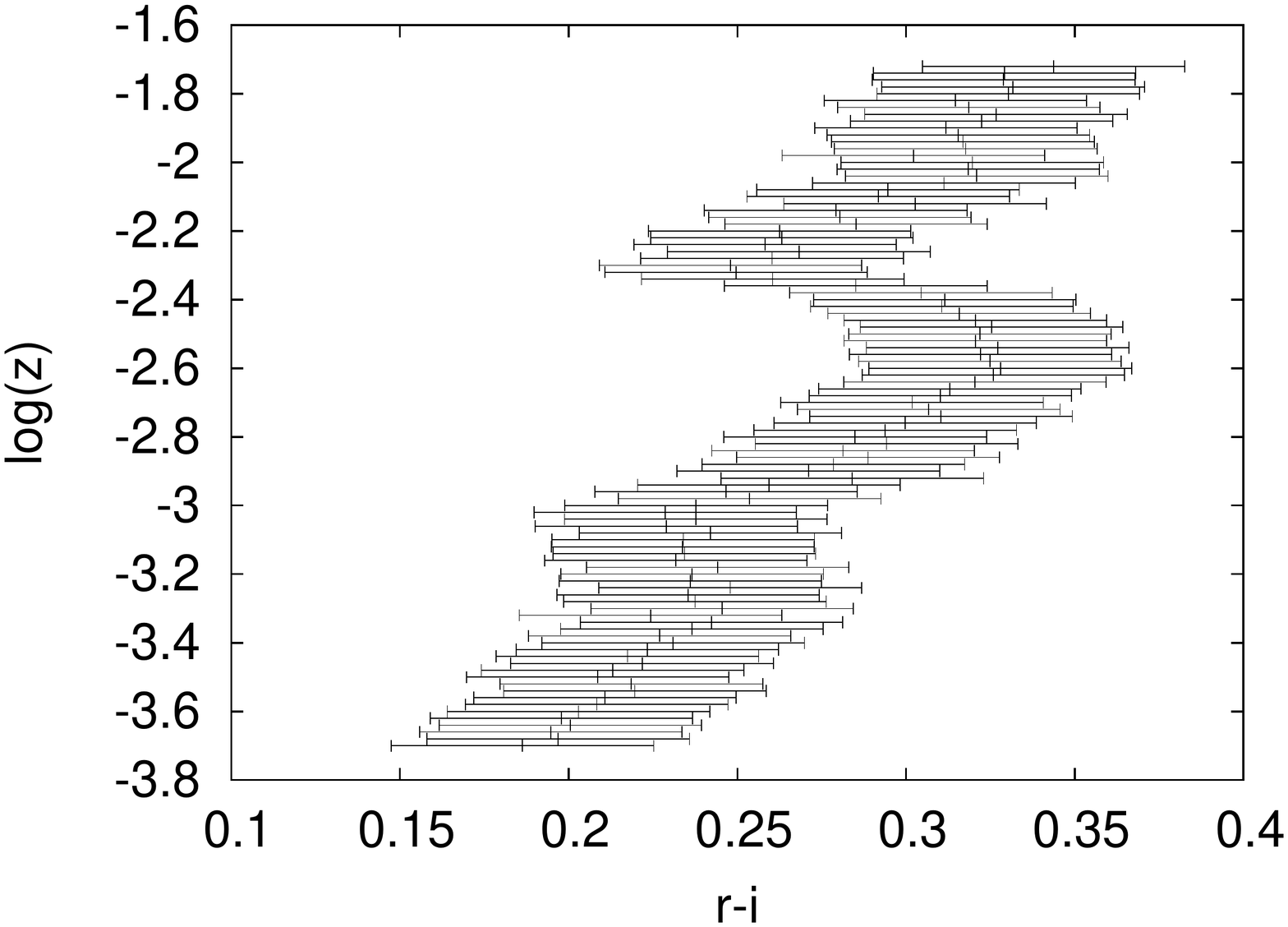}
  \includegraphics[width=0.45\textwidth]{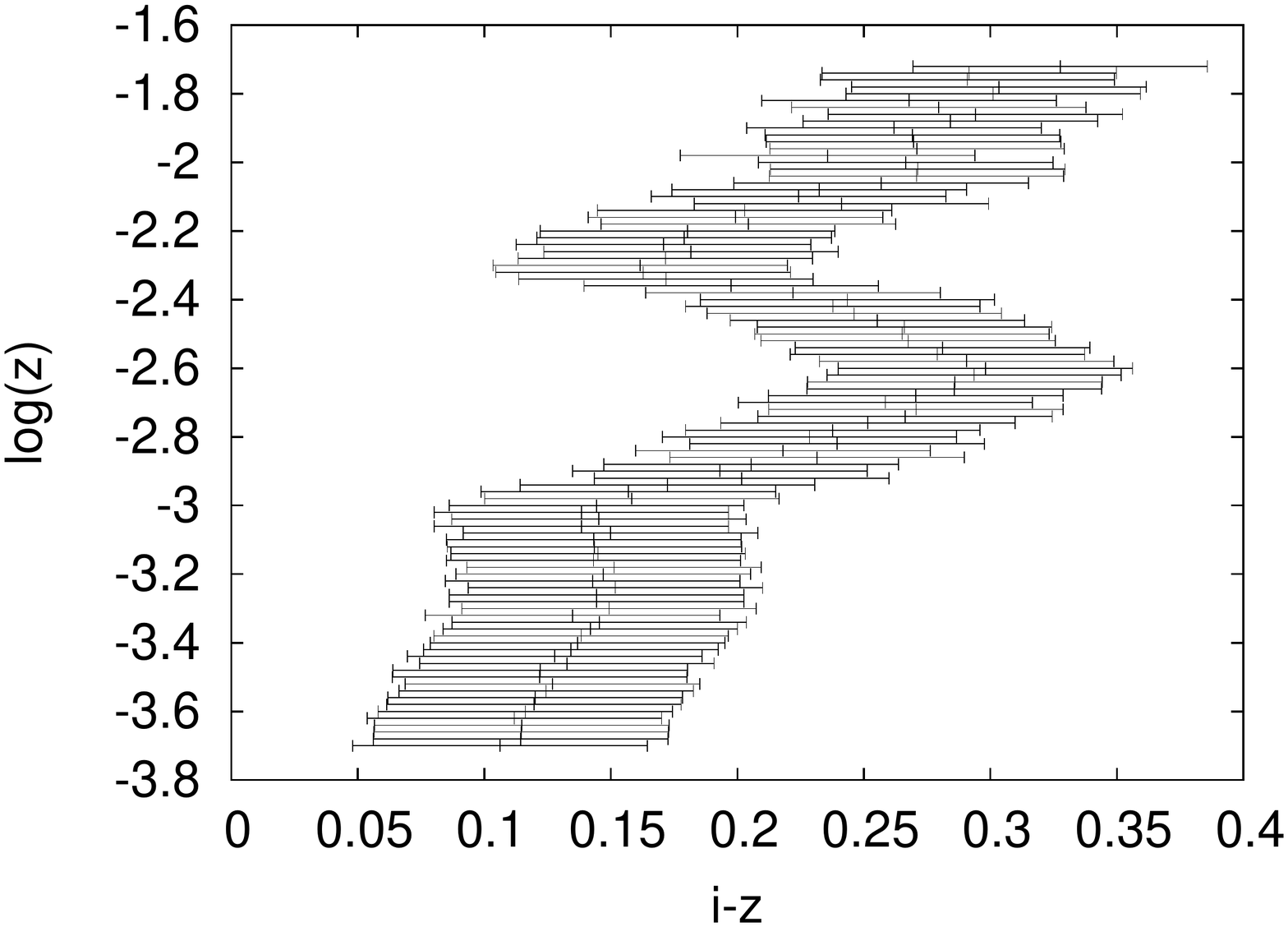}
  \caption{All combinations of ugriz luminosities to colors vs. metallicity. Each data point is the color at an individual metallicity averaged over cluster ages larger than 4~Gyr.}
  \label{fig:zcol}
\end{figure*}

 Fig. \ref{fig:zcol} shows the average color over all time steps with cluster ages above 4~Gyr and their standard deviation from all 100 models for four combinations of ugriz colors. Each data point represents such an average for an entire cluster with the denoted metallicity. As shown in section \ref{sec:evolution}, in the optical colors computed from u, g and r there is little development with time after an age of 4~Gyr, whereas the greater spread in UV colors is due to larger changes in i and z.

A \lq wiggle\rq\, of different strength and shape is obvious in all colors, the feature being more prominent with the average wavelength of both wavelength bands increasing. The reason for this is stellar evolution on the Giant Branch. At high metallicities close to solar metallicity the first and second Giant Branches have tips at similar color. Quickly intermediate stars between the two tips appear (Red Horizontal Branch, RHB), which vanish at metallicities around $\log(z)=-2.36$. This process is accompanied by an elongation of the second Giant Branch towards red color with decreasing metallicity, which accelerates as soon as the RHB stars have vanished. The tip of the second Giant Branch in principle moves linearly from left to right with decreasing metallicity, but together with the Red Horizontal Branch stars disappearing at that point performs a movement back and forth around $\log(z)\approx -2.5$, causing the wiggle feature in Fig. \ref{fig:zcol}. These stellar evolution processes participating in the RGB phase transition should therefore lead to a bimodality in color. 

\noindent GloMAGE (\textbf{Glo}bular Cluster \textbf{M}etallicity and \textbf{AGE}) reads an input file containing an observed cluster's designation together with the corresponding ugriz absolute luminosities with errors and determines probable metallicities. The expected metallicity value is found using Bayes' theorem

\[
  P(z_i|c) = \frac{P(c|z_i)P(z_i)}{\sum\limits_{j=1}^n P(c|z_j)P(z_j)},
\]

where $z_i$ is the $i^\mathrm{th}$ metallicity and $c$ is an observed color. In frequentist statistics the underlying assumption is the law of large numbers approximating a probability distribution by a Gaussian distribution which is determined by a mean $\mu$ and a standard deviation $\sigma$. The general idea of Bayesian statistics is that this is not necessarily the case. Instead, prior knowledge of the problem is used to determine an empirical probability distribution function (pdf). In both cases the expectation value of the distribution can be calculated from the usual definition 

\[
  \operatorname{E}[X] = \sum_{i=1}^\infty x_i\, P(x_i),
\]

\noindent assuming that the series converges absolutely. In the case of all physical problems considered in this paper this is fulfilled because of physical values being finite.

One of the strengths of the Bayesian approach to statistical analysis is that, in contrary to frequentist statistics breaking down at small sample sizes, Bayesian statistics leads to reliable results also in that case, at the cost of requiring prior knowledge of the distribution. Even though usually the prior is unknown to some extent, a sufficiently meaningful data sample leads to posteriors only marginally influenced by the choice of the prior. Therefore, the usual approach is to iterate the prior by using the posterior as a prior for a second repetition.

In the case of this paper, an observed integrated color is compared to $N_z$ models with different metallicities. As a prior one can choose a uniform distribution so that the probability of the model $m_i$ resulting in the observed color value $c$ is equal $1/N_z$ for all models. The difference between model value and observational value can be expressed in terms of the uncertainty of the simulation $\sigma$ by $\Sigma=\left|\frac{m_i-c}{\sigma}\right|$. This number represents an approximation for the probability that the observed value actually does not come from that model: One finds

\[
  P(c|m_i)=2-2\Phi(\Sigma),
\]

\noindent where $\Phi(\Sigma)$ is the cumulative distribution function $\Phi(\Sigma)=\frac{1}{\sqrt{2\pi}} \int\limits_{-\infty}^{\Sigma} e^{ -\frac{1}{2}t^2 }\, dt$ of the Gaussian distribution. If $N_c$ such observables are combined to form a joint posterior distribution (in the ugriz case, $N_c=10$), Bayes' theorem leads to

\[
  P(m_i|c) = \prod\limits_{k}P(m_i|c_k)=\frac{1-\Phi(\Sigma_k)}{N_m-\sum\limits_{j}\Phi(\Sigma_{jk})}.
  \label{eq:bayesmult}
\]

The general workflow of GloMAGE for each observed cluster is as follows:
\begin{enumerate}
  \item Read observational color data for all clusters,
  \item For the $i^\mathrm{th}$ cluster in the input list: Compare each observed color to the colors of each model in the catalog. For computing metallicities from observed colors Bayesian statistics is used. The deviation of the observed value to the catalog data is quantified by $P(c|z_i)=2-2\Phi(\sigma)$. Multiply the probability density functions for each color to obtain a total pdf,
  \item Normalize the probability density function to unit total probability,
  \item Compute the expectation value $E(c)=\sum\limits_{i}P(z_i|c)z_i$,
  \item Fit a Gaussian distribution function to the pdf to obtain the full width at half maximum as an uncertainty,
  \item Check obtained metallicities with observational data.
\end{enumerate}

All uncertainties by means of the evaluation technique introduce errors to the metallicity. As sources for uncertainties several factors have to be considered:
\begin{itemize}
  \item[-] The statistical scatter of the Monte Carlo simulation.
  \item[-] The initial particle number.
  \item[-] The value of the initial tidal radius.
  \item[-] The initial binary fraction.
  \item[-] The distance to the cluster system.
  \item[-] The zero-points of the underlying magnitude system.
  \item[-] Observational errors.
\end{itemize}

\noindent The statistical scatter of the simulation increases up to 0.05 magnitudes in z band and implies an error in color of up to 0.06. An error in color of this magnitude causes an error in terms of metallicity of in most cases not more than $\Delta(\log{z})\approx 0.3$.

So far most of the simulations run used initial particle numbers of $50,000$. A different size of the system will certainly affect luminosities, but as preliminary tests have shown color is unaffected.

A too low value of the initial tidal radius leads to more escaped stars and therefore to lower luminosity. In principle, color is also affected, but the change in color is insignificant in terms of the simulation and is neglected.

The binary fraction and the initial concentration do not affect the development of luminosity largely after an age of 4~Gyr assuming that binaries can not be resolved. The binary fraction, however, changes color (see section \ref{sec:evolution}). The largest difference in color is found between realizations with binary fractions of 0\% and non-zero binary fractions, because primordial binaries change the effective initial mass function of a cluster by doubling the mass of the object the simulation considers. The color shift introduced by binary fractions between 10\% and 80\% is about 0.1 magnitudes. The systematic effect on metallicity is similar to that of the statistical scatter but is smaller than $\Delta (\log{z})\approx 0.2$.

Uncertainties in the distance to the cluster system and the radius of that galaxy's halo introduce errors to luminosity, but, as they are systematic, not to color.

Also the uncertainty in determining the zero-points of the magnitude system adds to the error of the method, because it results in a systematic error in luminosity and therefore in age. As these uncertainties are smaller than 0.01 magnitudes, which is of the order of both observational and simulation errors, they can be neglected. Uncertainties caused by any other observational error have to be taken into account using data provided in the input file. Luminosities can be measured to high accuracy and errors usually do not exceed 0.1 magnitudes. This causes metallicity differences of up to $\Delta(\log{z})\approx 0.4$.

\section{Testing GloMAGE using M31 Data}
\label{sec:m31}

All types of galaxies seem to contain globular clusters and so does the Andromeda Galaxy, M31. First observed by Hubble in 1932, catalogs of the M31 globular cluster system have been compiled both using ground-based telescopes (e.g. van den Bergh in 1969, Huchra in 1982 or Christian and Heasley in 1991) and the Hubble Space Telescope (e.g. Grillmair in 1996). The Revised Bologna Catalog by \citet{Galleti09a} is a classic source of globular cluster positions, luminosities, metallicities and, to some extent, ages. Several hundred clusters are known (\citet{Caldwell11}) and have been observed thoroughly in ugriz by \citet{Caldwell09}. \citet{Peacock10a} have performed photometry in the ugriz and K bands of 416 old and 156 young globular clusters in M31 from images of the SDSS survey.  Among others, \citet{Fan06}, \citet{Caldwell09} and \citet{Caldwell11} compared observations to Simple Stellar Populations.

\citet{Caldwell11} used the MMT Hectospec Multifibre Spectrograph with a resolution of 5~\r{A} in a wavelength range of between 3700~\r{A} and 9200~\r{A} to find the total metallicity of M31 globular clusters. From the spectra Lick indices are computed and transformed into cluster metallicities in [Fe/H] using calibration data obtained from Milky Way globular clusters. This assumes that globular clusters are coeval in both Milky Way and M31 and that both galaxies had similar star formation histories. These assumptions result in uncertainties not recoverable so far (\citet{Caldwell11}).

Two major problems of the method using Hectospec data to find a cluster's metallicity are a break in the conversion between Lick indices and [Fe/H] at $\approx -1.5$, which agrees well with results presented in section \ref{sec:glomage}, specifically shown in Fig. \ref{fig:zcol}, and insensitivity of Balmer line indices above $\mathrm{[Fe/H]}\approx -1$. Therefore, above values of $\mathrm{[Fe/H]}\approx -1$ catalog data obtained from Lick indices have to be treated cautiously and our results presented in the following may deviate from refrence catalog values for $\mathrm{[Fe/H]}>-1$.

The metallicity distribution of M31 is very different from that of the Milky Way, which gives rise to doubts concerning the assumption of similar star formation histories in the observed and the galactic cluster system. While the Milky Way's globular cluster system has a bimodal metallicity distribution function with peaks at $\mathrm{[Fe/H]}=-1.5$ and $\mathrm{[Fe/H]}=-0.7$, most authors (e.g. \citet{Kim07}, \citet{Caldwell09} and \citet{Caldwell11}) agree that the metallicity distribution of the M31 system is either unimodal with a peak at $\mathrm{[Fe/H]}=-0.9$, but no clear Gaussian or trimodal metallicity distribution function.

\begin{figure}
  \centering
  \includegraphics[width=0.5\textwidth]{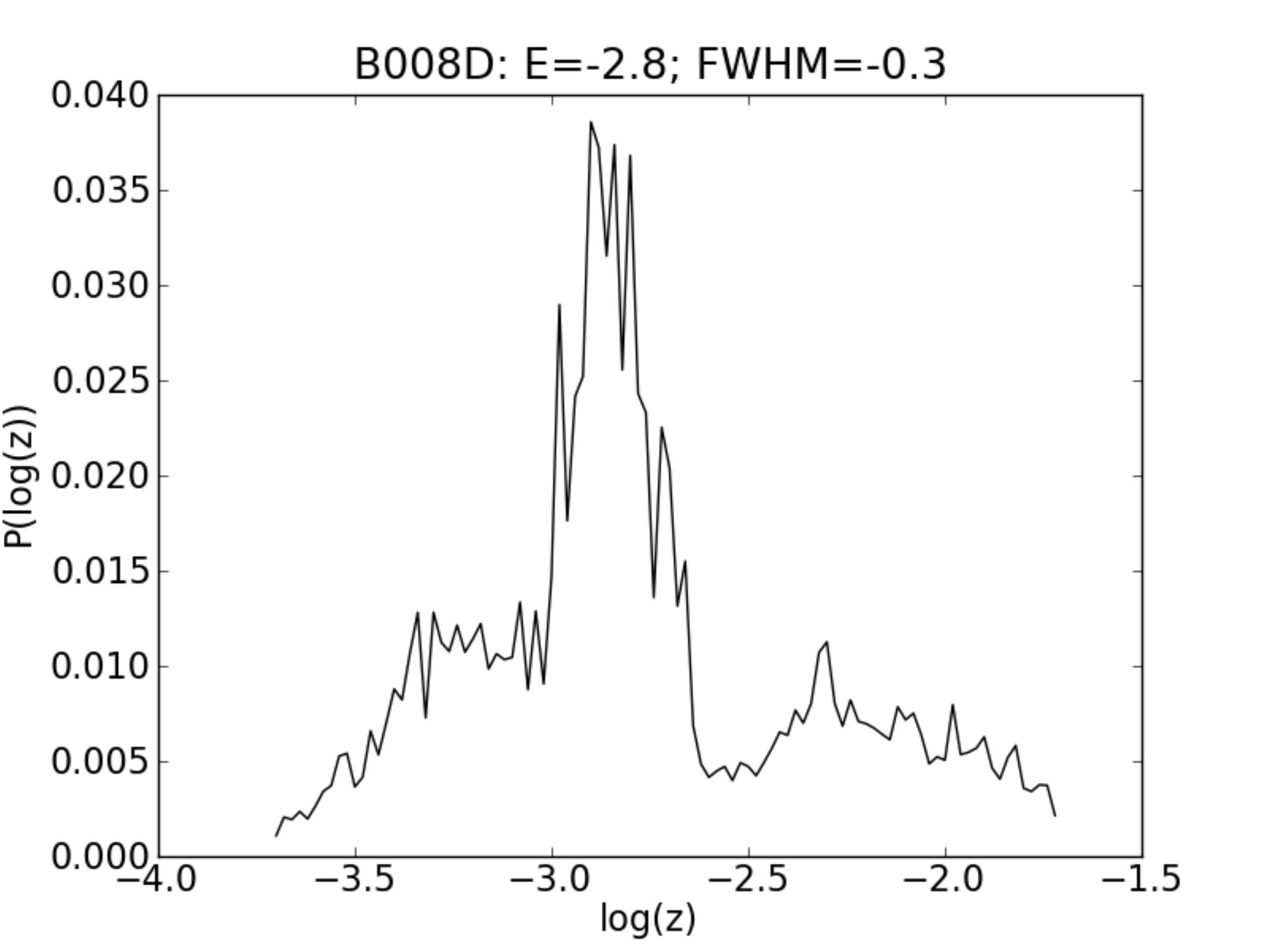}
  \caption{Probability density function for metallicity of the M31 globular cluster B008D.}
  \label{fig:pdf}
\end{figure}

\begin{figure}
  \centering
  \includegraphics[width=0.5\textwidth]{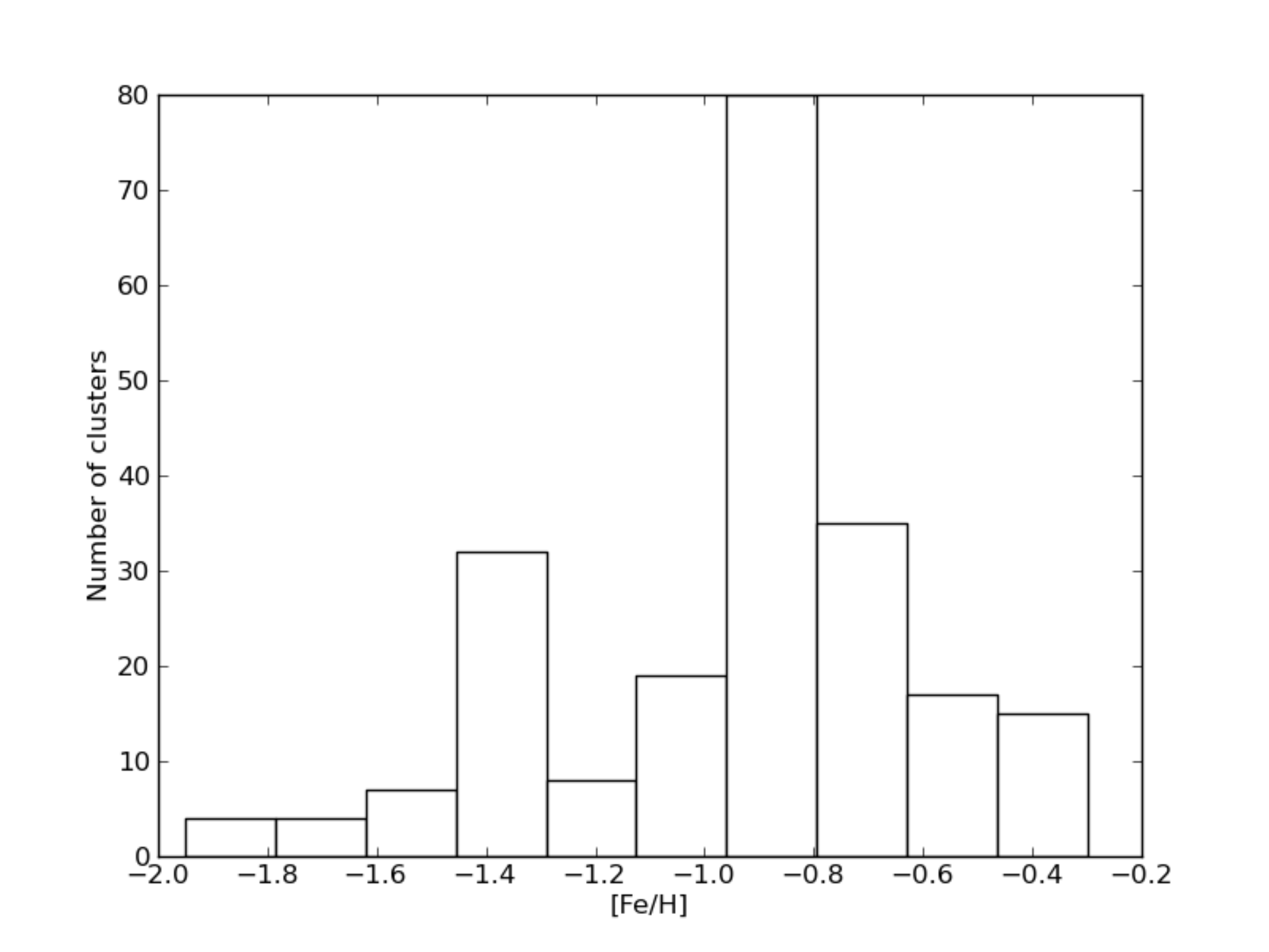}
  \caption{Metallicity histogram of M31 globular clusters analyzed with GloMAGE.}
  \label{fig:m31zdist}
\end{figure}

For comparisons with simulations we used the luminosity data provided by \citet{Peacock10a}, because they used SDSS data which is compatible with the AB system of the ugriz filters used for the compilation of the GloMAGE catalog. These luminosities were converted into absolute luminosities using the distance modulus found by \citet{Graeve81}. For reference a part of the catalog found using GloMAGE is shown in Tab. \ref{tab:zcatm31}. The catalog lists the logarithm of the metallicity with error and a reference metallicity taken from \citet{Caldwell11}. If a value is not found, it is denoted as \textit{99.0}, because machine-readable catalog data supplied by SIMBAD requires columns of data to contain identical data types in all rows. It is noteworthy, that the metallicities found here are systematically higher than those found by \citet{Caldwell11}, which can be explained by them measuring [Fe/H] and not the total metallicity and them using Lick indices to determine metallicity. If alpha enhancement of about 0.2 dex is considered, the agreement improves drastically. In the latter case, only four out of 221 clusters are found to have metallicities differing from reference data by more than 3 sigma. The catalog data of three of these four clusters indicates metallicities of below $\log{z}=-3.7$ and therefore outside of the applicable range of the method. Therefore, for more than 95\% of the clusters correct metallicity data was found. The metallicity distribution function found using GloMAGE is shown in Fig. \ref{fig:m31zdist} and displays a result relatively similar to the observational distribution. As the metallicity range above $\mathrm{[Fe/H]}\approx -1$ is difficult to determine from Lick indices, any difference in that region has to be taken cautiously and the peak at $\mathrm{[Fe/H]}\approx -0.9$ may be real. 

\begin{table}
\centering
\caption{Part of the M31 globular cluster metallicity catalog compared to \citet{Caldwell11} (SSP). The columns show the cluster ID taken from \citet{Caldwell11}, $\log{z}$ and its error determined by GloMAGE and reference metallicities with errors from the Caldwell catalog. Errors are $1\sigma$ errors in both cases.}
\begin{tabular}[c]{c|cccc}
  \hline
  \multicolumn{5}{c}{Part of the M31 GC [Fe/H] catalog}\\
  \hline
  ID & $\log{z}$ & $\Delta\log{z}$ & $\log{z}_\mathrm{ref}$ & $\Delta\log{z}_\mathrm{ref}$\\
  \hline
  \input{cat_m31_part.txt}
  \hline
\end{tabular}
\label{tab:zcatm31}
\end{table}

\section{Explaining the Color Bimodality}
\label{sec:bimodality}

There are 10 distinct color combinations of ugriz luminosities. As coeval cluster groups do not develop at the same time but in a time span of around 1~Gyr (see for instance \citet{Marin-Franch09} who show that relative ages of galactic cluster populations have a root-mean-square scatter of about 10\%, meaning up to 1~Gyr for 10~Gyr old cluster systems), it is statistically equivalent to consider $N$ clusters with different random seeds (i.e. different realizations) and to consider only one realization, but at $N$ time steps in a range of 1~Gyr. Even though the subsequent time steps for a single cluster are not statistically independent, they can be treated as if they were, because the intrinsic scatter of the Monte Carlo method is about as large as the scatter encountered in a single cluster's evolution. For snapshot times the time spans between 1 and 2~Gyr, 4 and 5~Gyr and 11 and 12~Gyr respectively have been chosen, so that the initial state of the color distribution, a \lq short term\rq\, development in early stages and the outcome after a long time span can be compared.

Because the bimodality is thought to originate from the metallicity distribution, four different metallicity distributions are studied. 

\begin{enumerate}
  \item[D1:] The entire range of all metallicities between $2\cdot 10^{-4}$ and $2\cdot 10^{-2}$ (uniform distribution), 
  \item[D2:] the lower half between $2\cdot 10^{-4}$ and $2\cdot 10^{-3}$ (uniform distribution),
  \item[D3:] the high-metallicity half between $2\cdot 10^{-3}$ and $2\cdot 10^{-2}$  (uniform distribution) and 
  \item[D4:] two extremal ranges between $2\cdot 10^{-4}$ and $6\cdot 10^{-4}$, and $9\cdot 10^{-3}$ and $2\cdot 10^{-2}$  (two uniform distributions to simulate a bimodal distribution). 
\end{enumerate}

\begin{figure*}
  \includegraphics[width=0.45\textwidth]{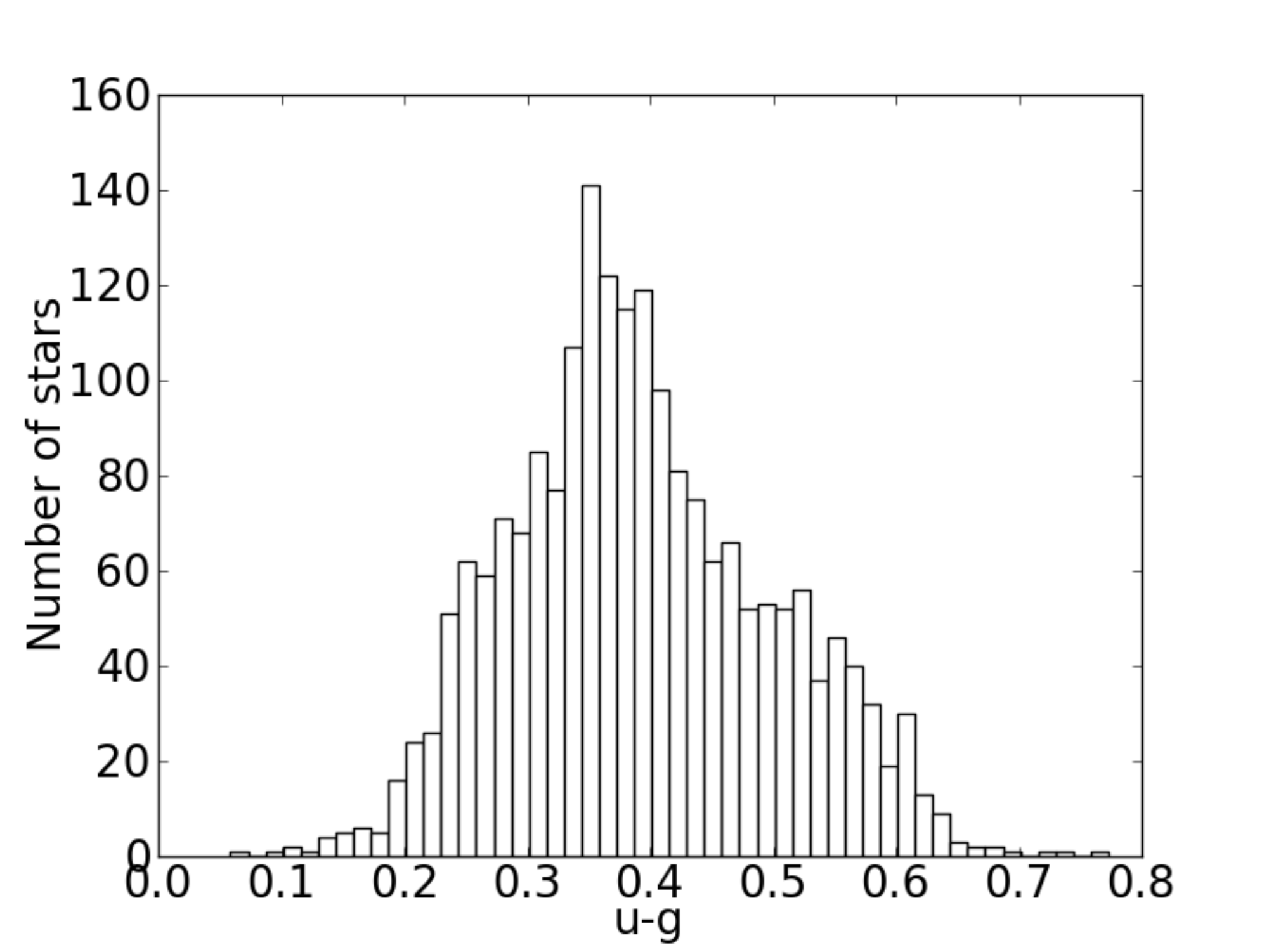}
  \includegraphics[width=0.45\textwidth]{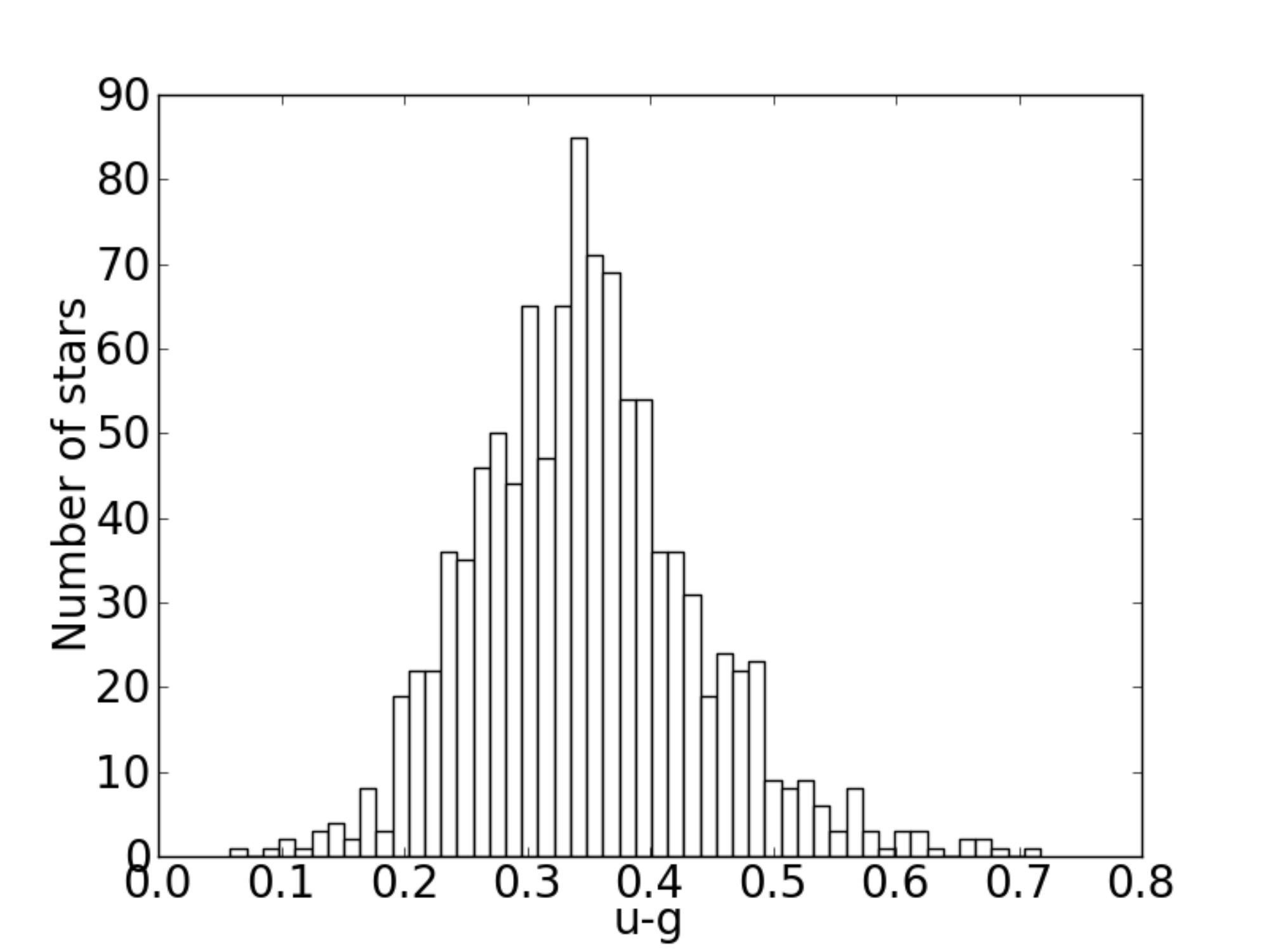}\\
  \includegraphics[width=0.45\textwidth]{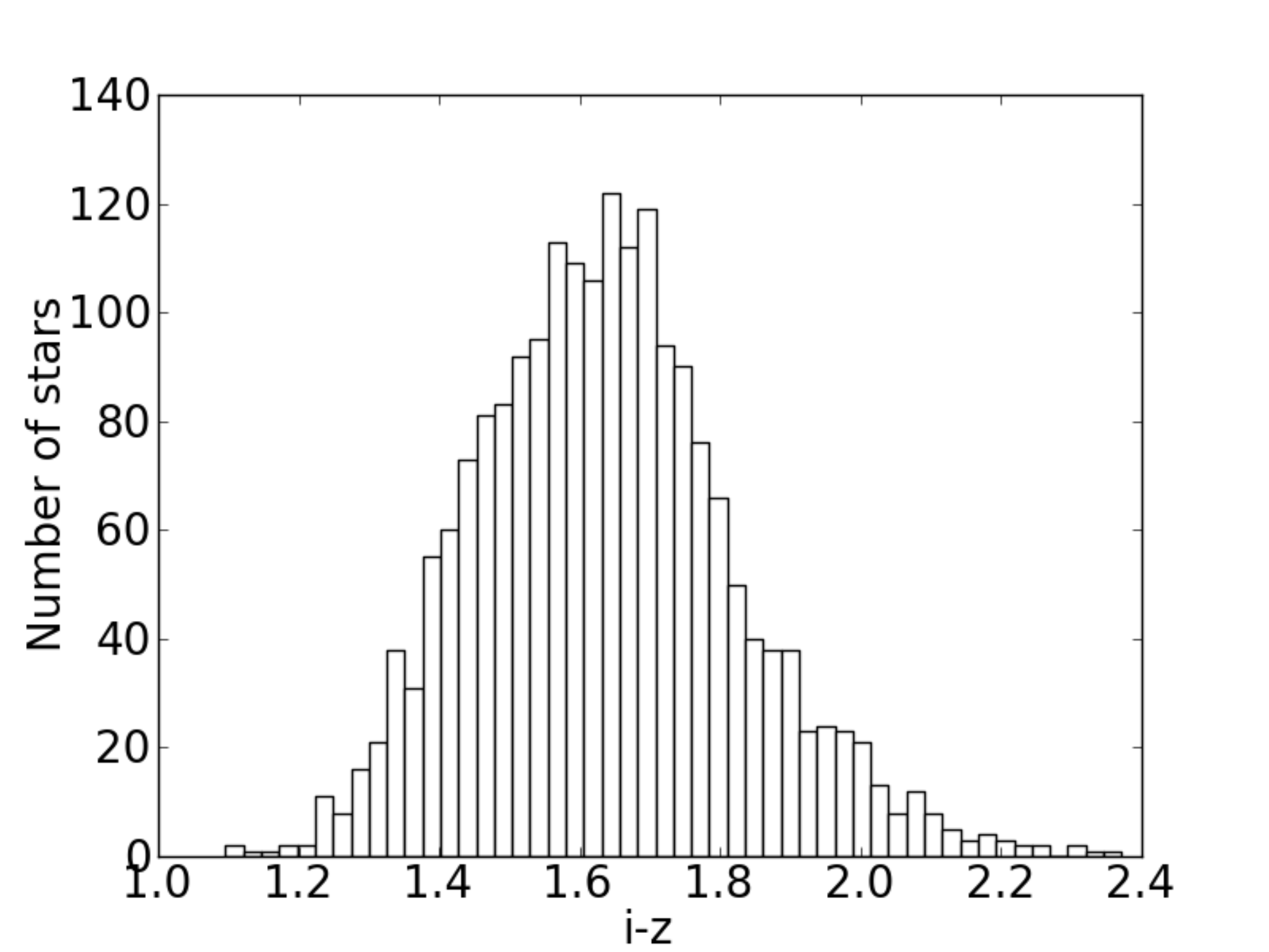}
  \includegraphics[width=0.45\textwidth]{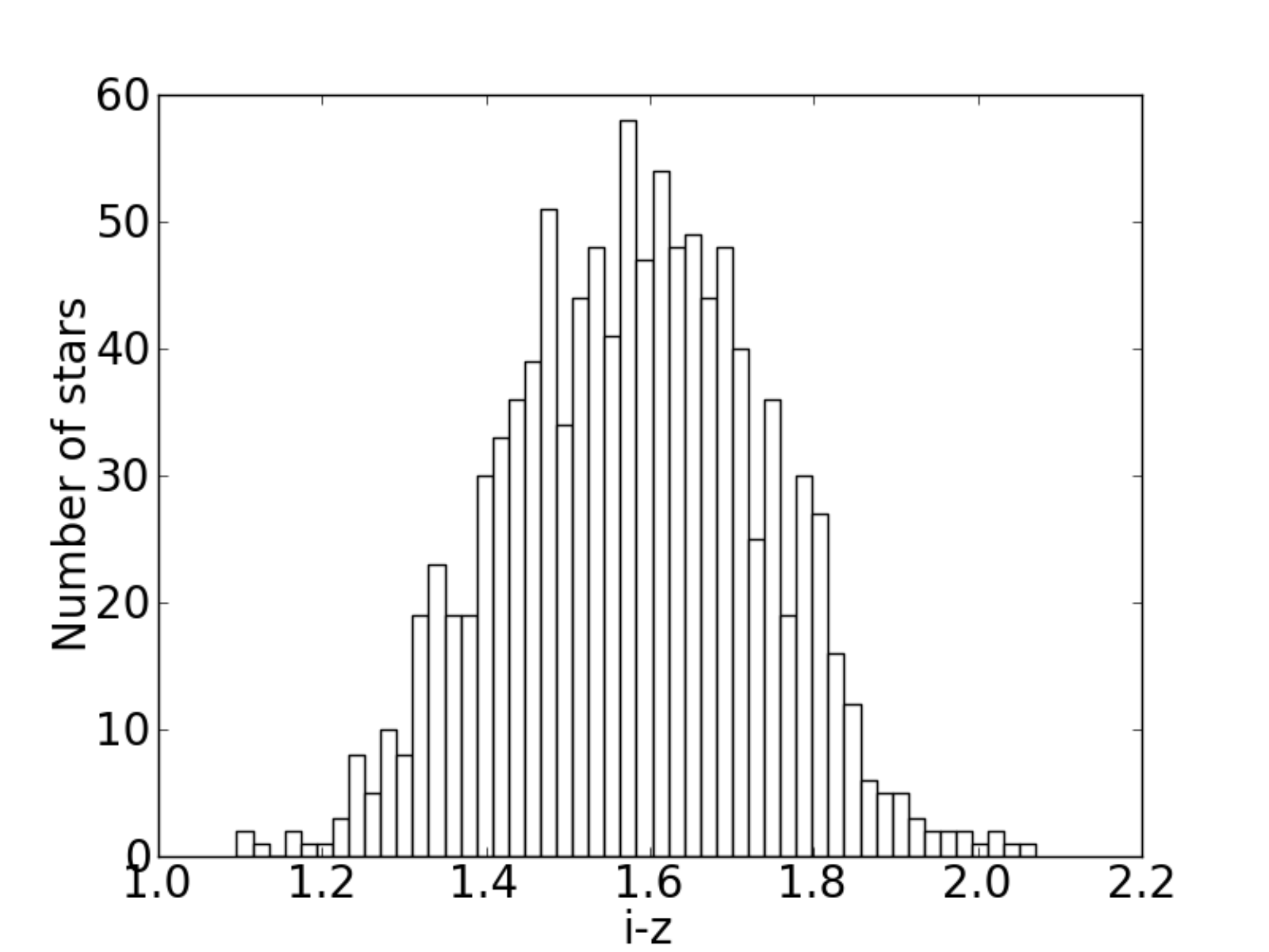}\\
  \includegraphics[width=0.45\textwidth]{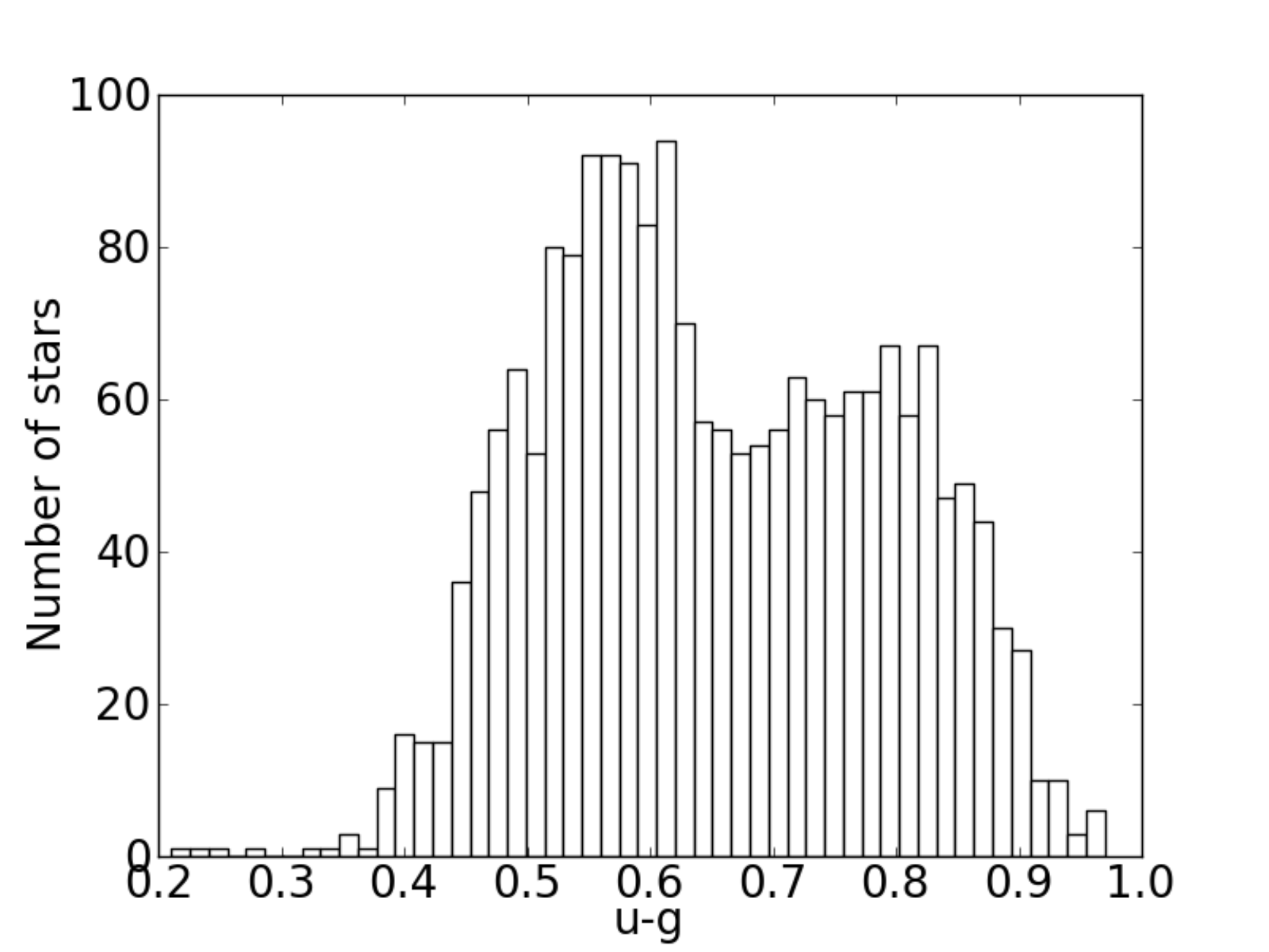}
  \includegraphics[width=0.45\textwidth]{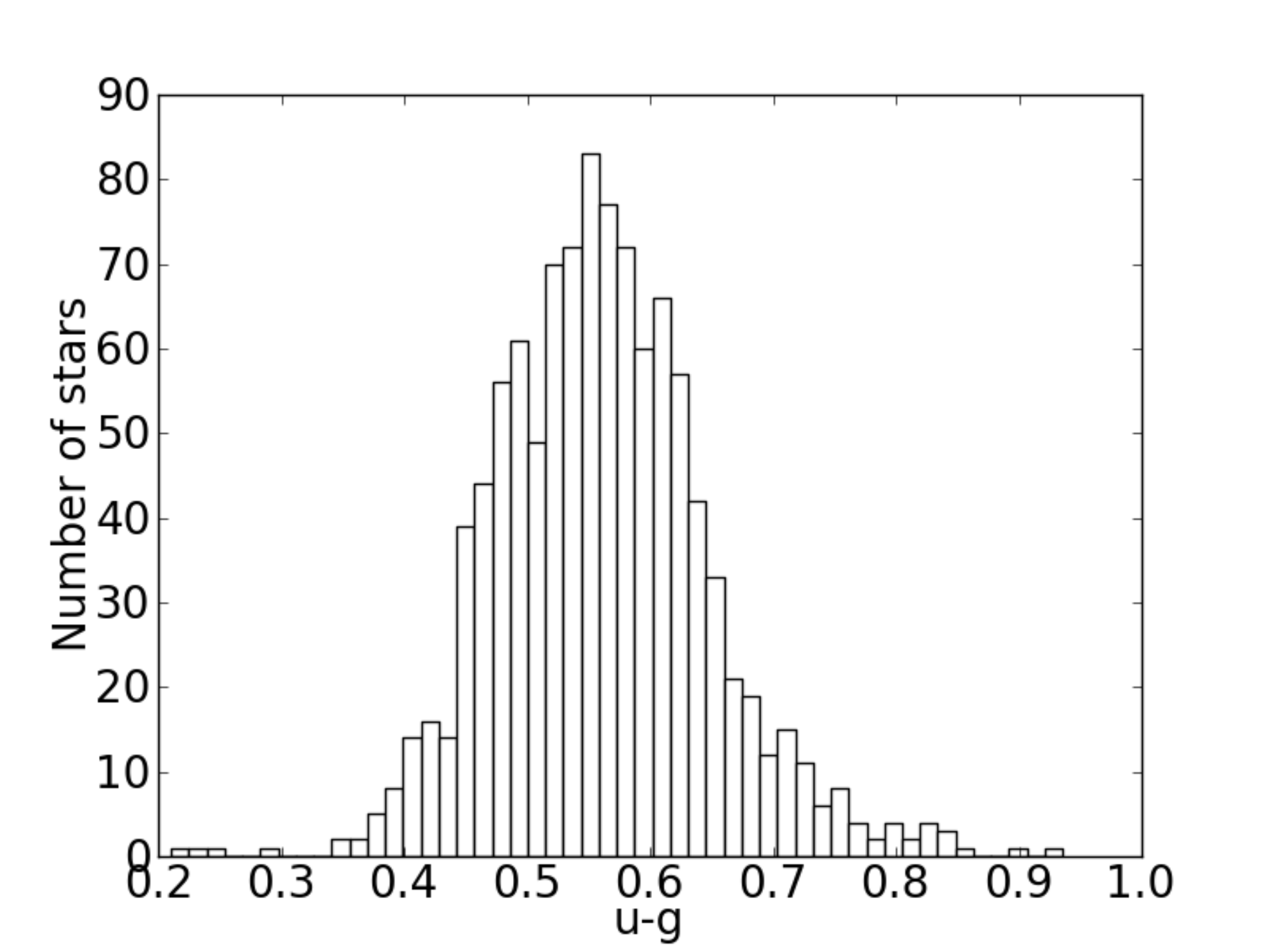}\\
  \includegraphics[width=0.45\textwidth]{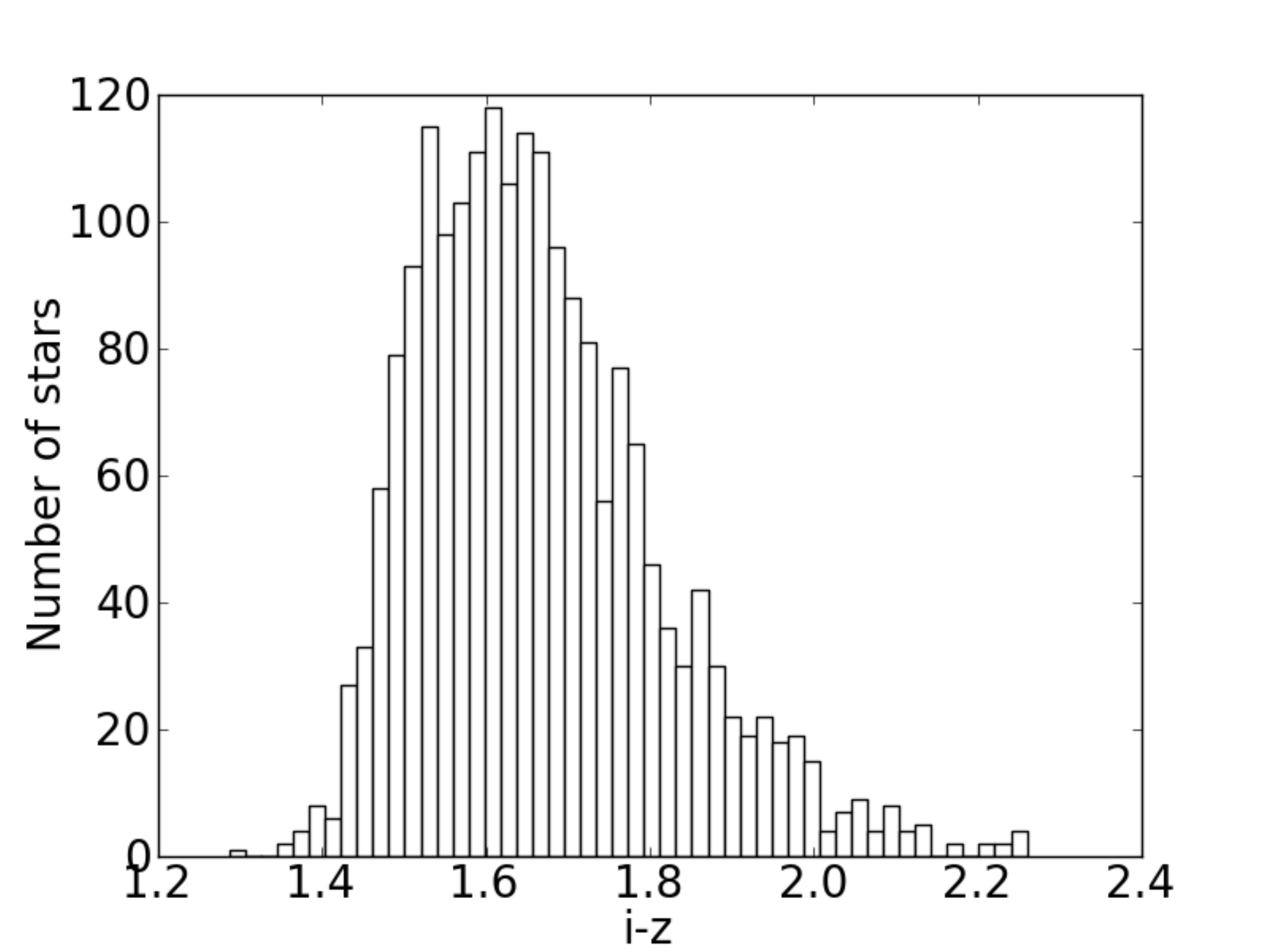}
  \includegraphics[width=0.45\textwidth]{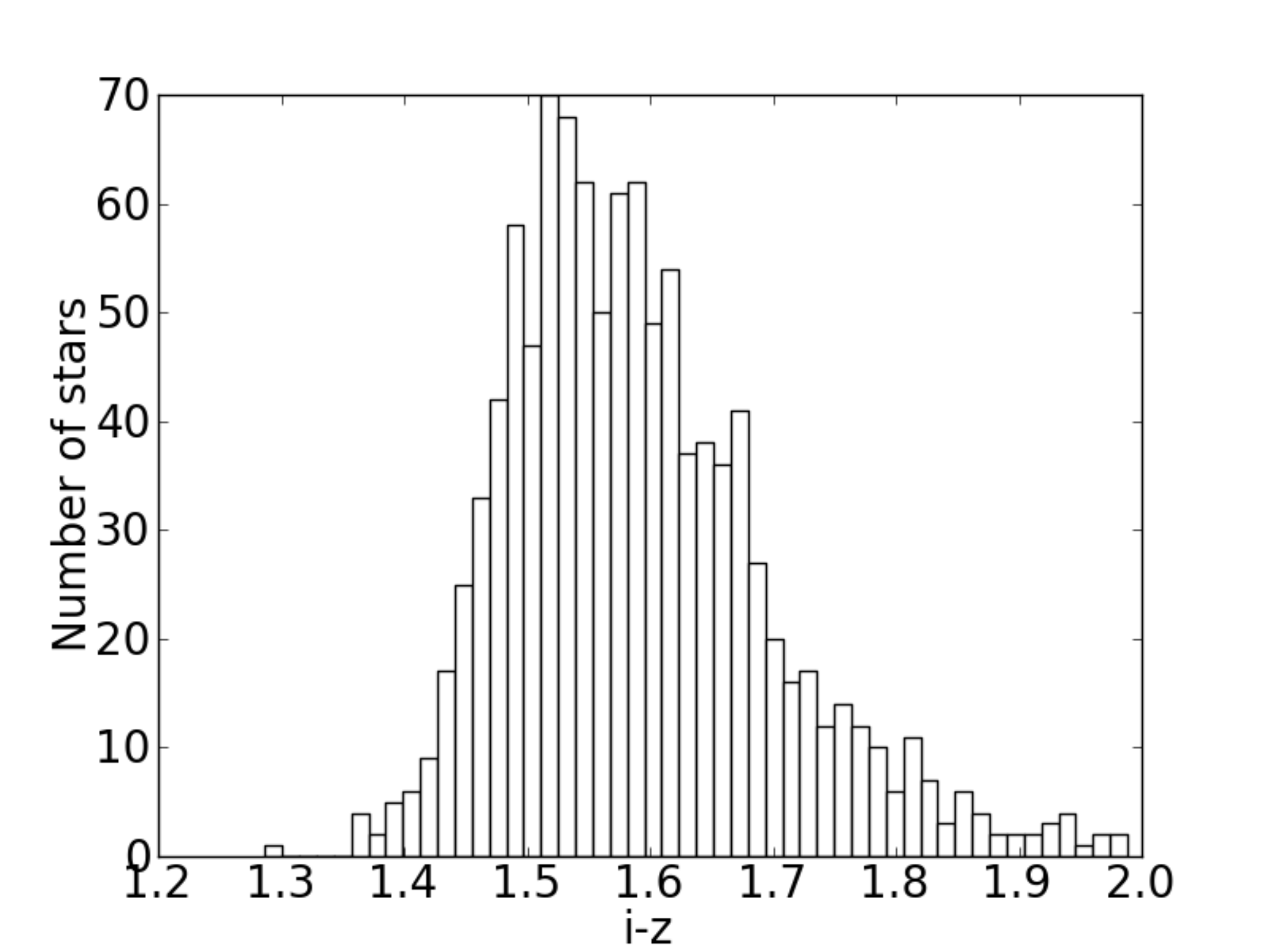}
  \caption{\label{fig:histcmp_a}Color histograms for distributions D1 (left column) and D2 (right column). The top two rows correspond to cluster ages of 2~Gyr (first line: u-g, second line: i-z), the bottom two to an age of 12~Gyr (first line: u-g, second line: i-z).}
\end{figure*}

\begin{figure*}
  \includegraphics[width=0.45\textwidth]{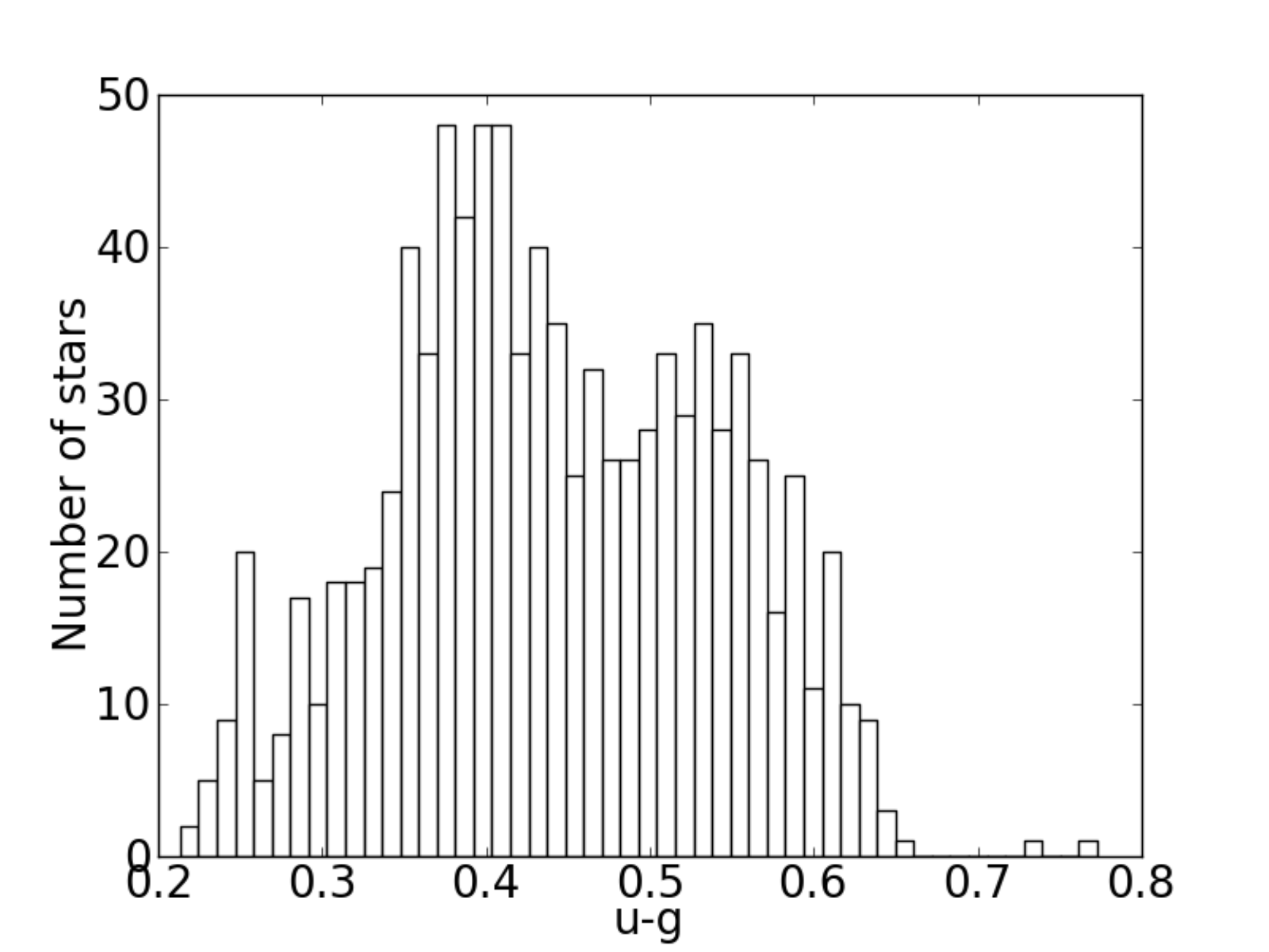}
  \includegraphics[width=0.45\textwidth]{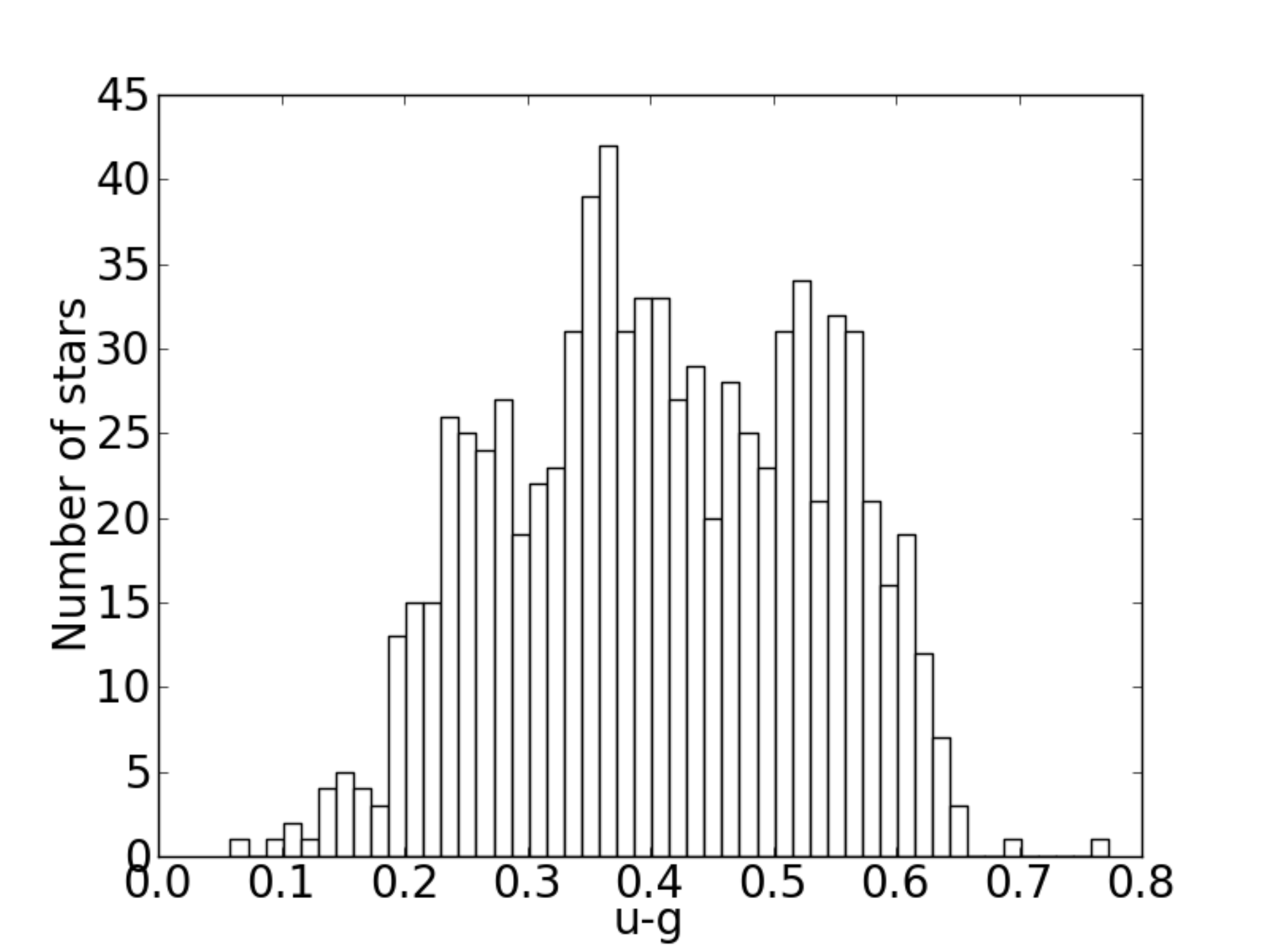}\\
  \includegraphics[width=0.45\textwidth]{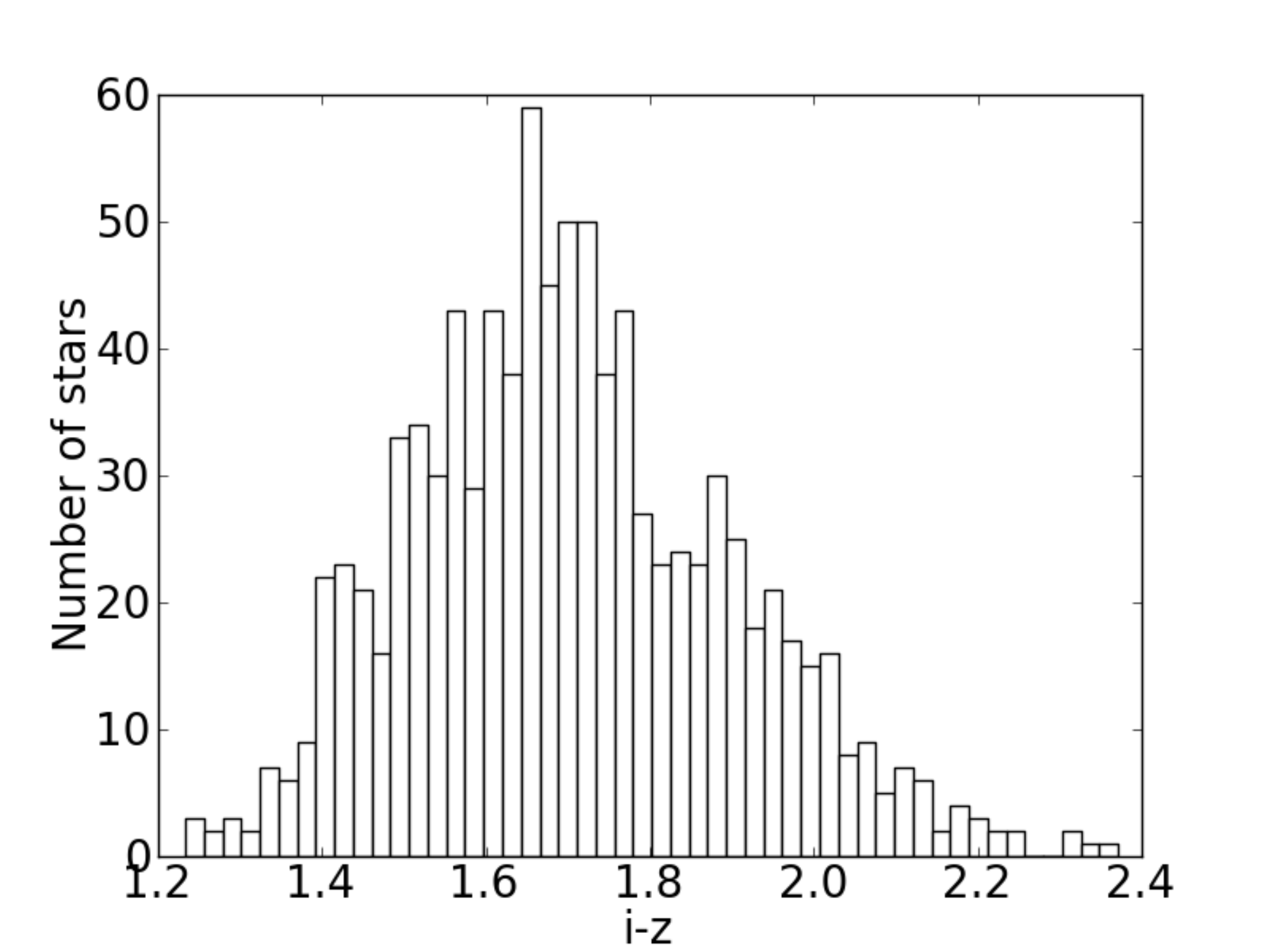}
  \includegraphics[width=0.45\textwidth]{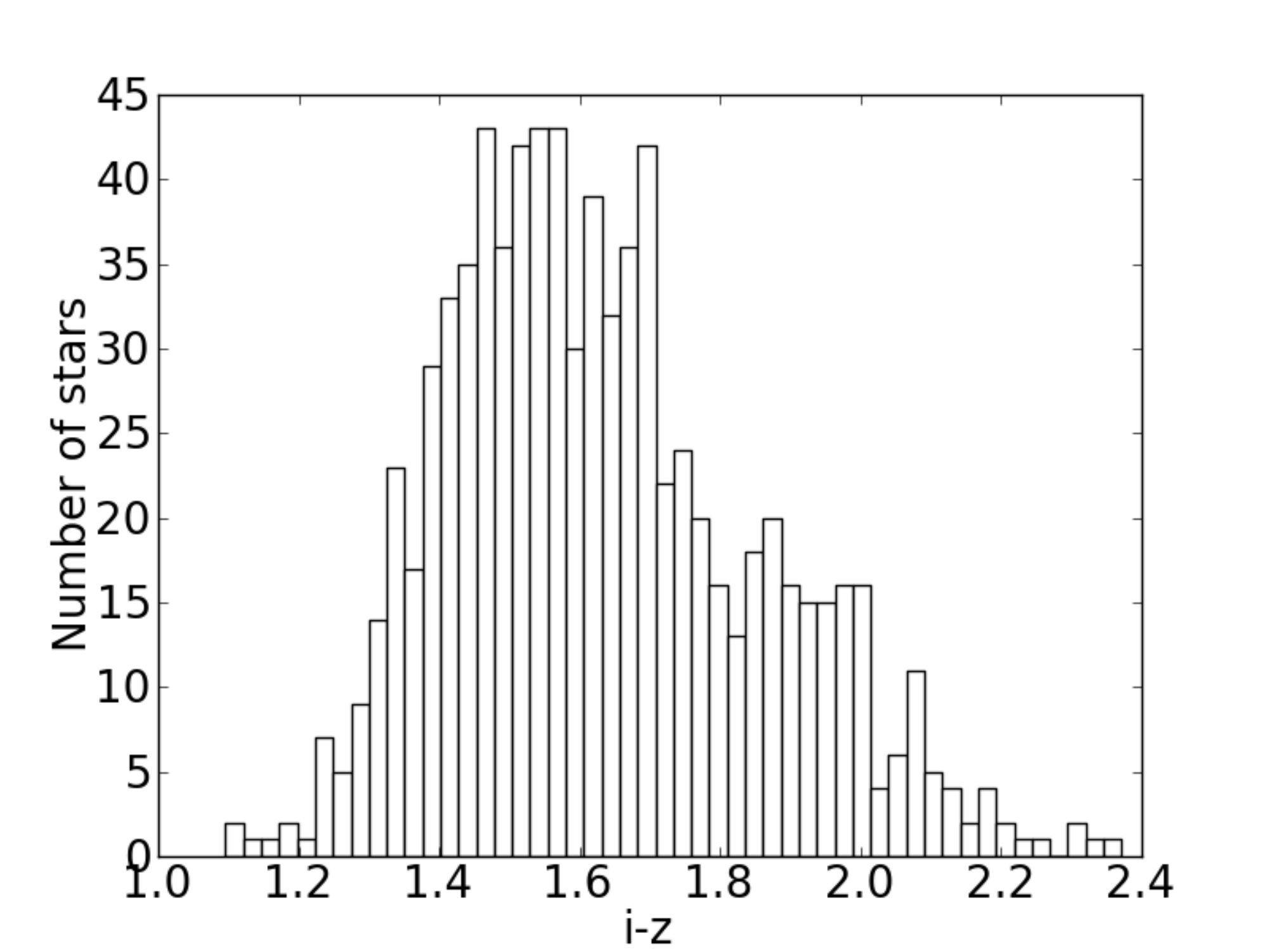}\\
  \includegraphics[width=0.45\textwidth]{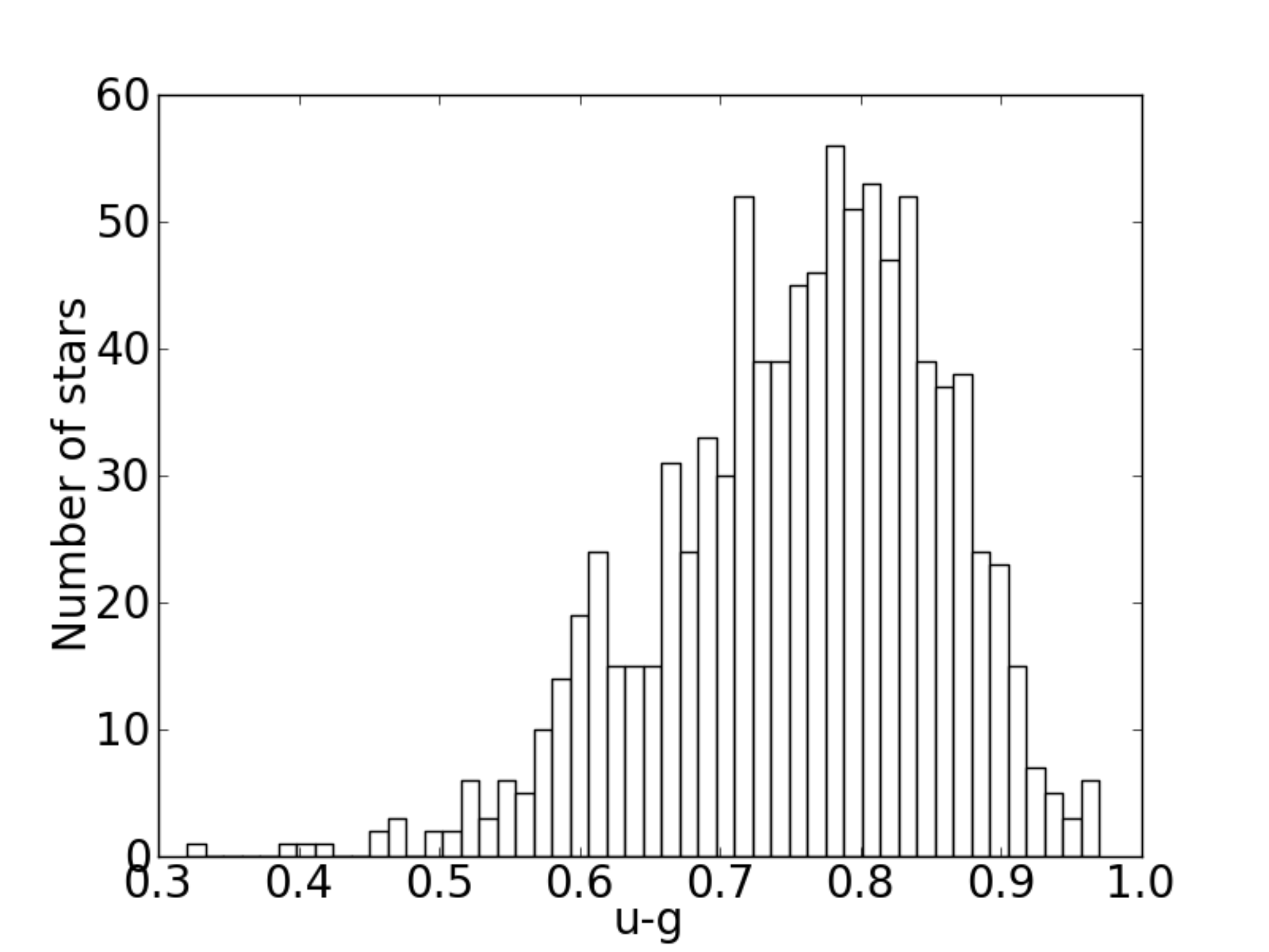}
  \includegraphics[width=0.45\textwidth]{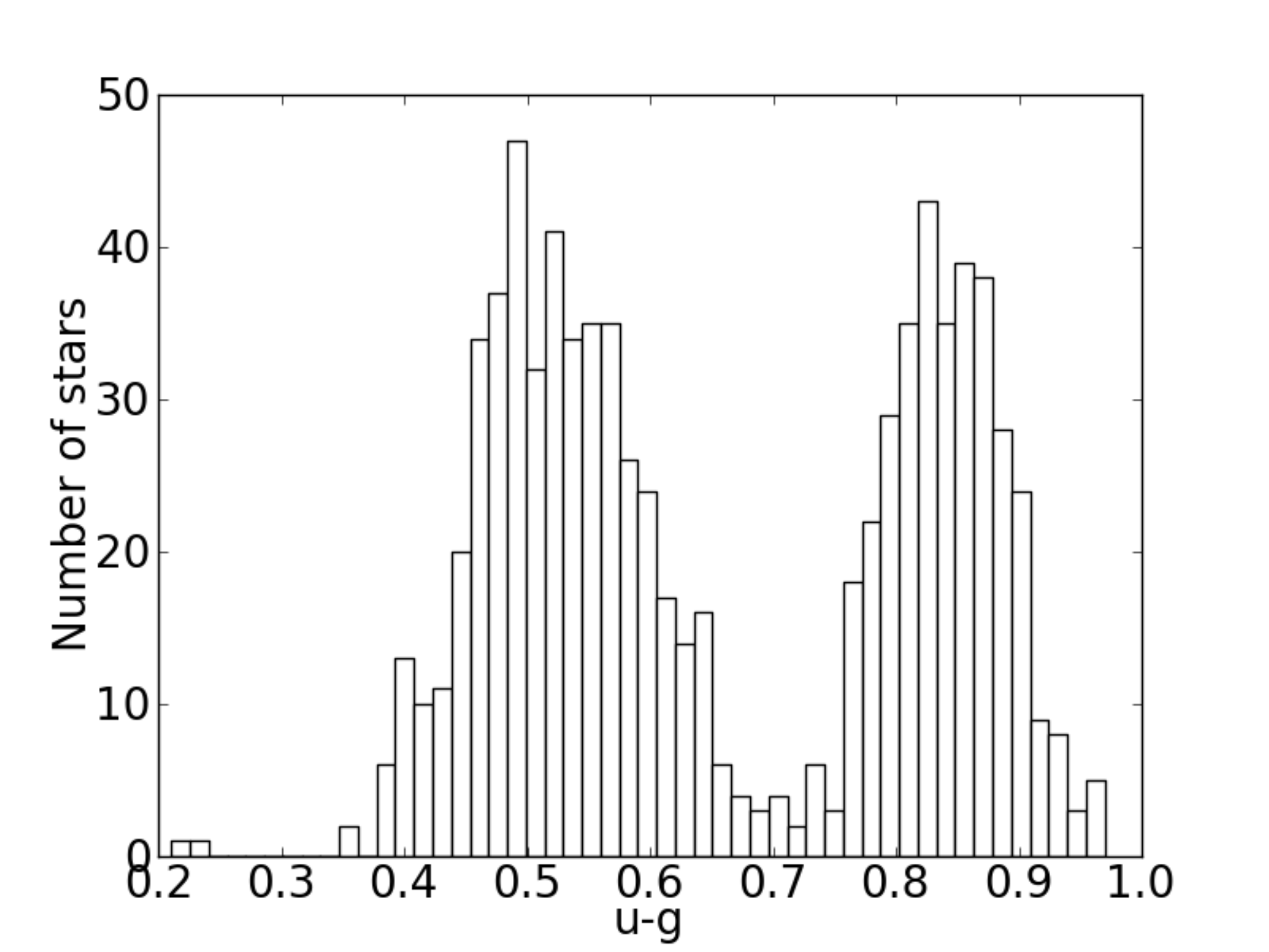}\\
  \includegraphics[width=0.45\textwidth]{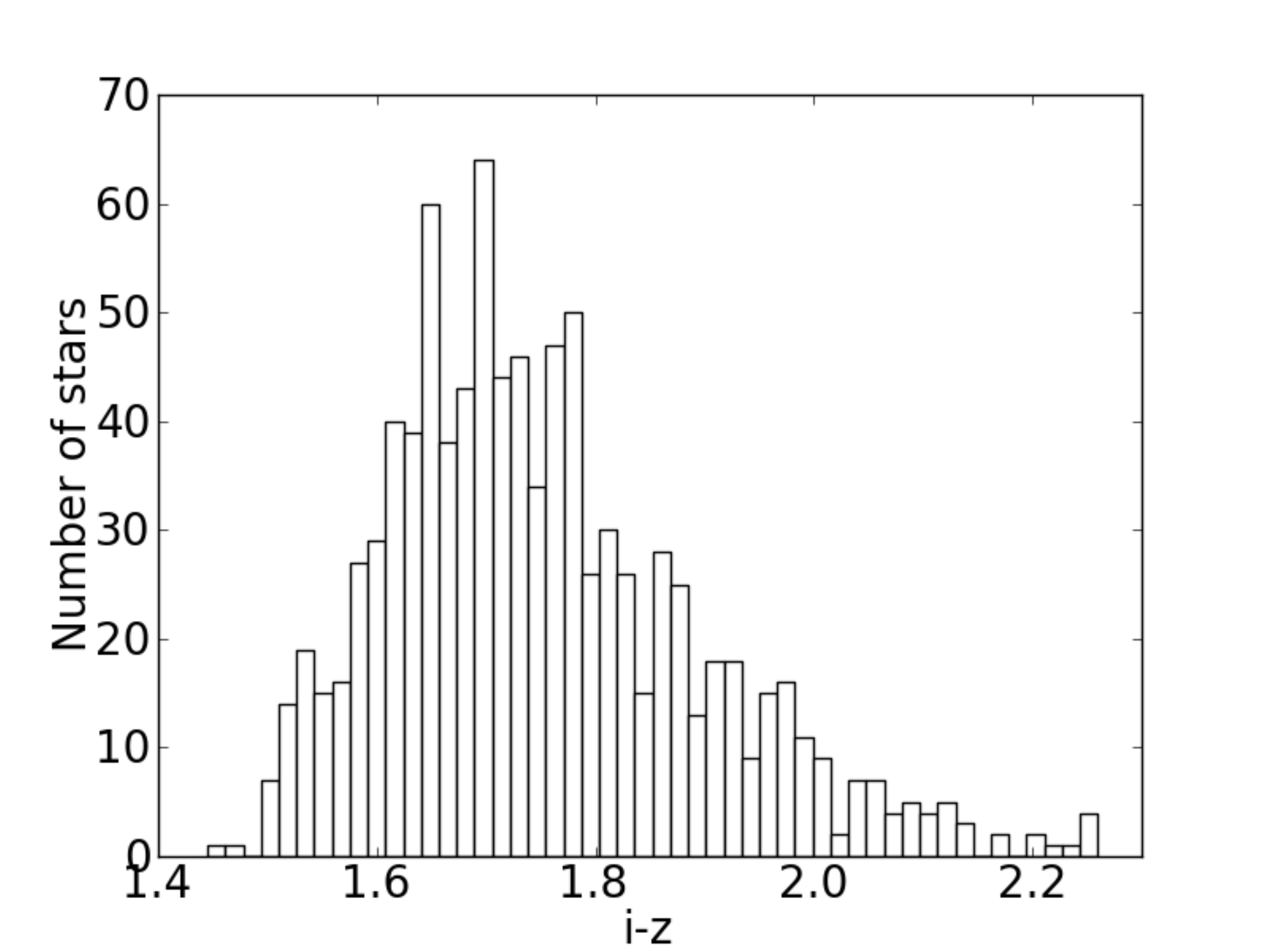}
  \includegraphics[width=0.45\textwidth]{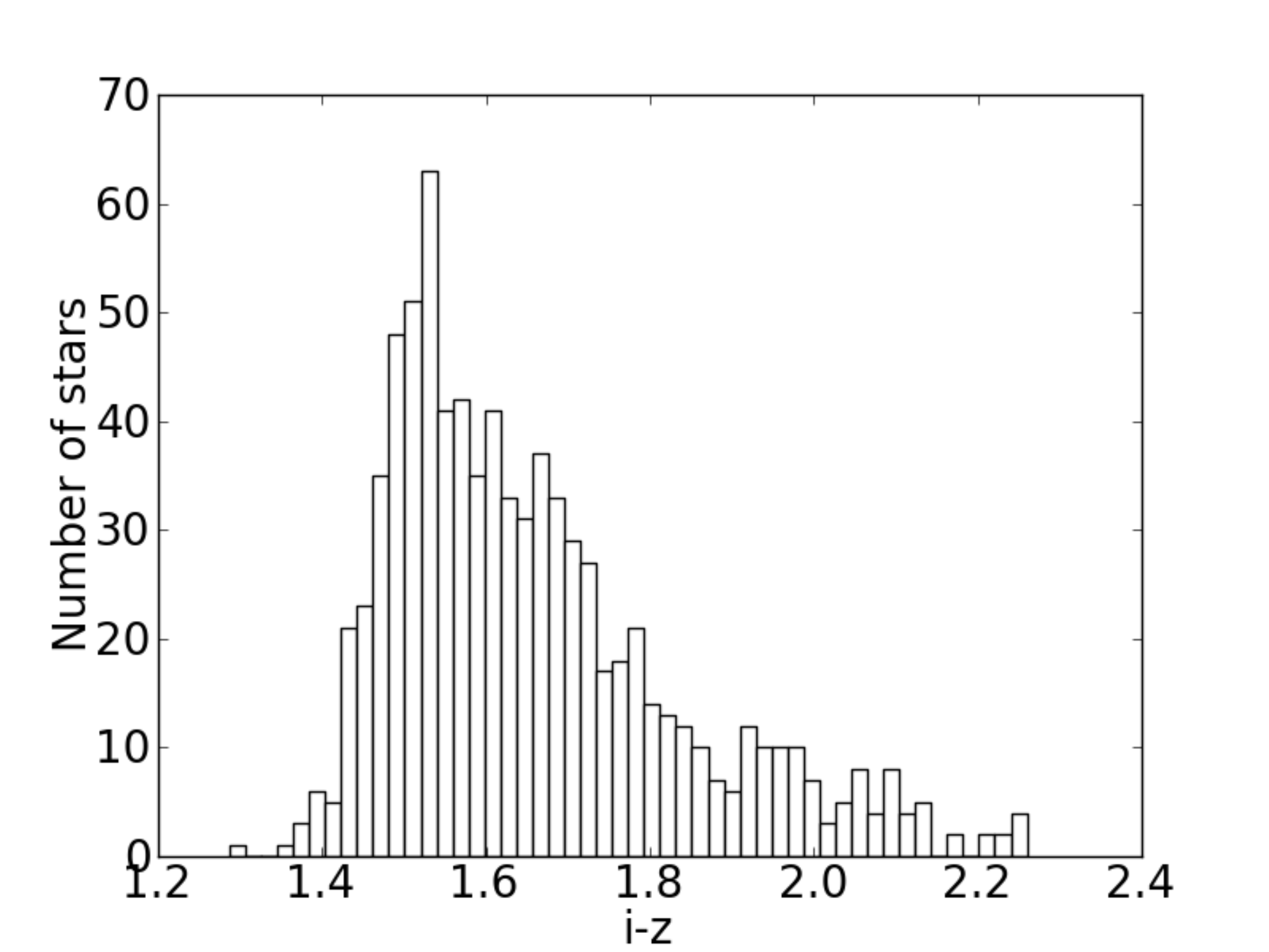}
  \caption{\label{fig:histcmp_b}Color histograms for distributions D3 (left column) and D4 (right column). The top two rows correspond to cluster ages of 2~Gyr (first line: u-g, second line: i-z), the bottom two to an age of 12~Gyr (first line: u-g, second line: i-z).}
\end{figure*}

Figs. \ref{fig:histcmp_a} and \ref{fig:histcmp_b} compare color histograms at an age of 2 and 12~Gyr. In principle, a bimodal distribution can develop in certain cases. Comparing the different time steps it is apparent that the colors subtracting from u and g band develop a stronger pronounced bimodality than the colors from i or z band, in which observational errors possibly blur remaining signatures of a bimodal distribution. This supports the observation that bimodalities are weaker in colors in the UV. The feature is most strongly developed in distributions 1 and 4, and weaker, but still apparent, in distribution 3. Distribution 2 displays a bimodality as well, but only in a very weak way and only at intermediate ages.

Distribution 1 does not display a bimodality, but the distribution broadens with time in all colors until at late times a weak bimodality appears in u-g, with the noteworthy feature that the blue peak is higher than the red peak. In distribution 2 there is no bimodality. The distribution fits a single Gaussian distribution at first, then it broadens and at late times restores its Gaussian form. The same is valid in distribution 3, but in this case the peak is shifted compared to that of distribution 2. Still, distributions 2 and 3 start out with similar distributions and the difference develops after several~Gyr as was expected from section \ref{sec:evolution}. Distribution 4 combines many of the features of distributions 2 and 3, but in a stricter sense, as the sub-distributions are confined tighter. Therefore, expecially in u-*, a very strong bimodality develops with time.

The conclusion from the previous section that the wiggle feature in Fig. \ref{fig:zcol} originates from stellar evolution, also results in the color distribution. It has to be kept in mind that fig \ref{fig:zcol} considers populations older than 4~Gyr. Of course a single color does not develop much after this age, but to remove uncertainties in conclusions it is safe to say that this is entirely valid after 5~Gyr. Therefore comparisons of this figure only allow interpretations of histograms for cluster ages of 11 to 12~Gyr (see figs. \ref{fig:histcmp_a} and \ref{fig:histcmp_b}). A very interesting observation concerns the development of the bimodality in distributions 2 and 3 and this emphasizes the influence of stellar evolution: Clusters with higher metallicities evolve faster than those with lower metallicity, so that the non-gaussian form of the near-UV colors in distribution 3 develops earlier than that in distribution 2. Both of these observations are combined both in distributions 1 and 4, so that especially in distribution 4 a very strong bimodality is observable after a few~Gyr.

Especially the histogram of u-g for distribution 1 at 12~Gyr compares well to the observed bimodality of the Virgo galaxy cluster globular cluster system shown in the bottom right figure of Fig. \ref{fig:harris10_02}, whereas the distribution shown in the top right figure of Fig. \ref{fig:harris10_02} compares better to a histogram from distribution 2. The distinctiveness of an observed bimodality therefore cannot be seen as only marker of metallicity distribution of a coeval cluster set. Groups of clusters with a bimodal color distribution are older than a few~Gyr, and younger sets of clusters should not display such a bimodality. Especially the results from distribution 4 could indicate that a very strong bimodality is caused by two episodes of star (cluster) formation, as late time distribution evolution is dominated by the low-metallicity contribution. Different ranges of metallicities included into the computation of the histograms lead to different distinctiveness of the bimodality. The non-uniqueness of the metallicity-color relation leads to more or less prominent bimodalities in different colors.

Therefore, several statements can be made:
\begin{enumerate}
  \item The bimodality is an effect of a nonlinearity in the color-metallicity relation.
  \item The \lq life-time\rq\, of the bimodality depends on metallicity and therefore on stellar evolution processes.
  \item Colors including only the UV show a weaker bimodality than those subtracting from visual bands.
  \item Cluster sets with a distinct bimodality are in principle older than those with only a weak bimodal distribution.
  \item A bimodal color distribution of coeval clusters implies (amongst others) a bimodal metallicity distribution, but a unimodal color distribution does not imply a unimodal metallicity distribution.
  \item A non-coeval cluster set can show a color bimodality with a unimodal metallicity distribution.
\end{enumerate}

\section{Application to the Virgo Cluster GC System}
\label{sec:vir}

The Virgo galaxy cluster is an extensively studied region of the local universe close to the Milky Way. \citet{Mei07} find that its center is located at a distance of about 17 Mpc and according to \citet{Cote04} it has more than 2000 member galaxies. Several objects of the Virgo cluster were already known to Messier (including catalog entries 4, 58, 59, 60, 61, 84 to 91, 98, 99 and 100) and by now the Virgo Cluster Catalog VCC contains 2096 entries covering about 140 deg$^2$.

\begin{figure}
  \centering
  \includegraphics[width=0.5\textwidth]{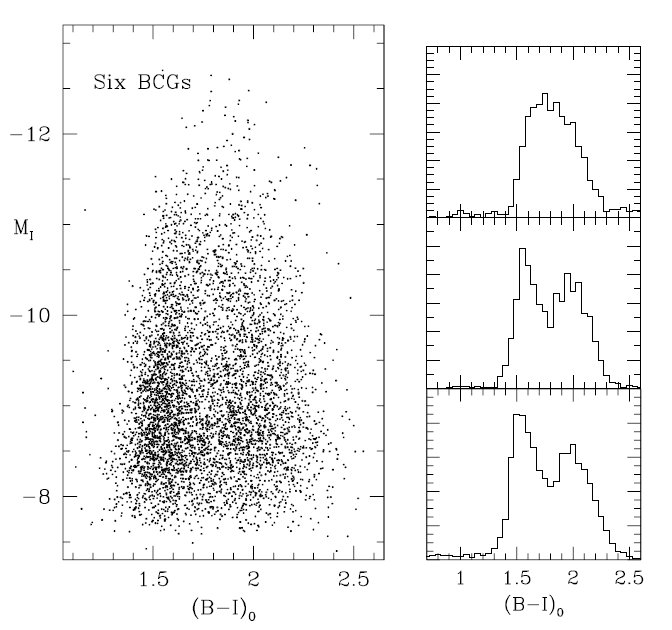}
  \caption{The color-magnitude distribution of 5500 globular clusters in six giant elliptical galaxies. The histograms show the color distributions for (from top to bottom) $M_I <-10.4$, $-10.4<M_I<-9.4$ and $-9.4<M_I<-8.4$. Taken from \citet{Harris10a}.}
  \label{fig:harris10_02}
\end{figure}

The Virgo cluster is centered around the giant elliptical galaxy M87, which has a large population of globular clusters (more than 10,000) studied intensively by \citet{Harris06}, \citet{Harris09} and \citet{Harris10a}. Using the \citet{Maraston05} stellar population synthesis models, he finds a bimodal metallicity distribution with peaks centered at $\mathrm{[Fe/H]}\approx -1.5$ and $\mathrm{[Fe/H]}\approx -0.4$ (see Fig. \ref{fig:harris10_02}).

This bimodality though is removed if only the most luminous clusters are considered. \citet{Harris09} also find that the low-metallicity cluster system is located further outside in the M87 cluster system, whereas the high-metallicity subsystem occupies a more centered ares. \citet{Harris10a} furthermore finds, that the bluer subsystem is 2~Gyr younger than the redder subsystem (even though it has to be noted that he gives an age error of 2~Gyr). 

This agrees with the observations of \citet{Peng11}, who find that the color distribution of the inner blue clusters of giant elliptical galaxies in the Coma cluster has a mean color of $g-I=0.94$, while in the case of the outer blue clusters the mean color is $g-I=0.89$; the red clusters in the inner part have a mean color of $g-I=1.18$.

In addition, the Advanced Camera for Surveys (ACS) installed on the Hubble Space Telescope was used by \citet{Cote04} in the ACS Virgo Cluster Survey to cover the globular cluster population 100 early-type galaxies in g and z band. \citet{Jordan09b} fit King profiles to the point-spread function of each cluster candidate to find the integrated luminositities and cut the expected luminosity range so that only expected old clusters with metallicities in the range $-2.5 \lesssim\mathrm{[Fe/H]}\lesssim 0$ are compiled into a catalog.

\begin{figure}
  \centering
  \includegraphics[width=0.5\textwidth]{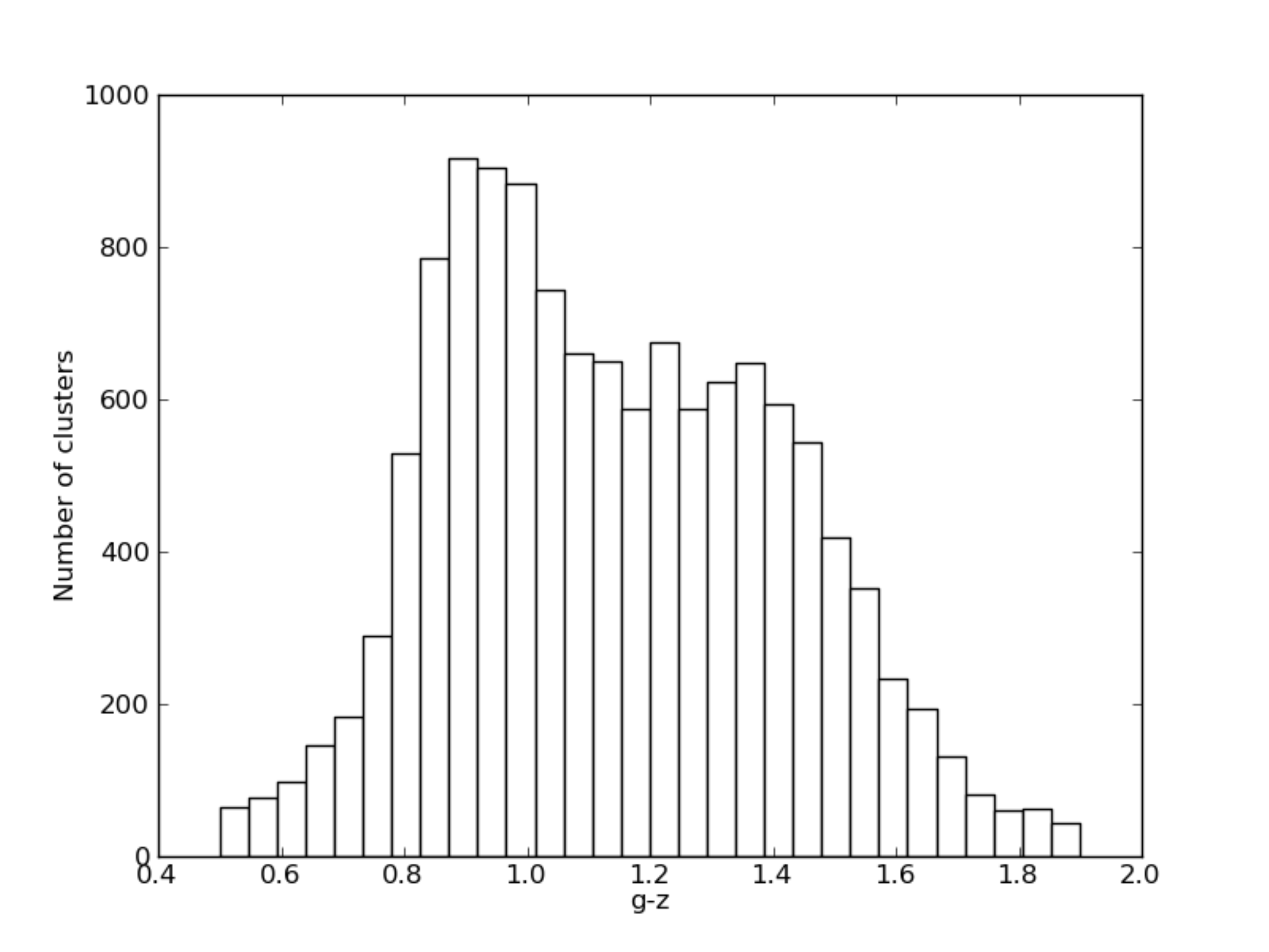}
  \caption{Virgo Cluster globular cluster color distribution in $g-z$ using data of the ACS Virgo Cluster Survey (\citet{Jordan09b}).}
  \label{fig:vircoldist}
\end{figure}

Fig. \ref{fig:vircoldist} shows the color distribution of globular clusters in the Virgo galaxy cluster in $g-z$ according to \citet{Jordan09b}. A clear bimodality is visible with peaks at colors $g-i\approx 0.9$ and $g-i\approx 1.35$ (the peak at $g-z=1.2$ can be interpreted as a relevant third mode, as will become clear below). The break between the two modes seems to be at approximately $g-z=1.2$.

\begin{table}
\centering
\caption{Part of the Virgo Cluster globular cluster metallicity and age catalog. The columns show the cluster ID and $\log{z}$ and its error determined by GloMAGE. Errors are $1\sigma$ errors. A value of 99.0 denotes a missing entry in the reference catalog. As a cluster ID right ascension and declination are given conform with the ACS Virgo Survey Catalog (\citet{Jordan09b}).}
\begin{tabular}[c]{ccc}
  \hline
  \multicolumn{3}{c}{Part of the Virgo Cluster GC [Fe/H] catalog}\\
  \hline
  ID & logz & logzerr\\\par
  \hline
  \input{cat_vir_part.txt}
  \hline
\end{tabular}
\label{tab:zcatvir}
\end{table}

\begin{figure}
  \centering
  \includegraphics[width=0.5\textwidth]{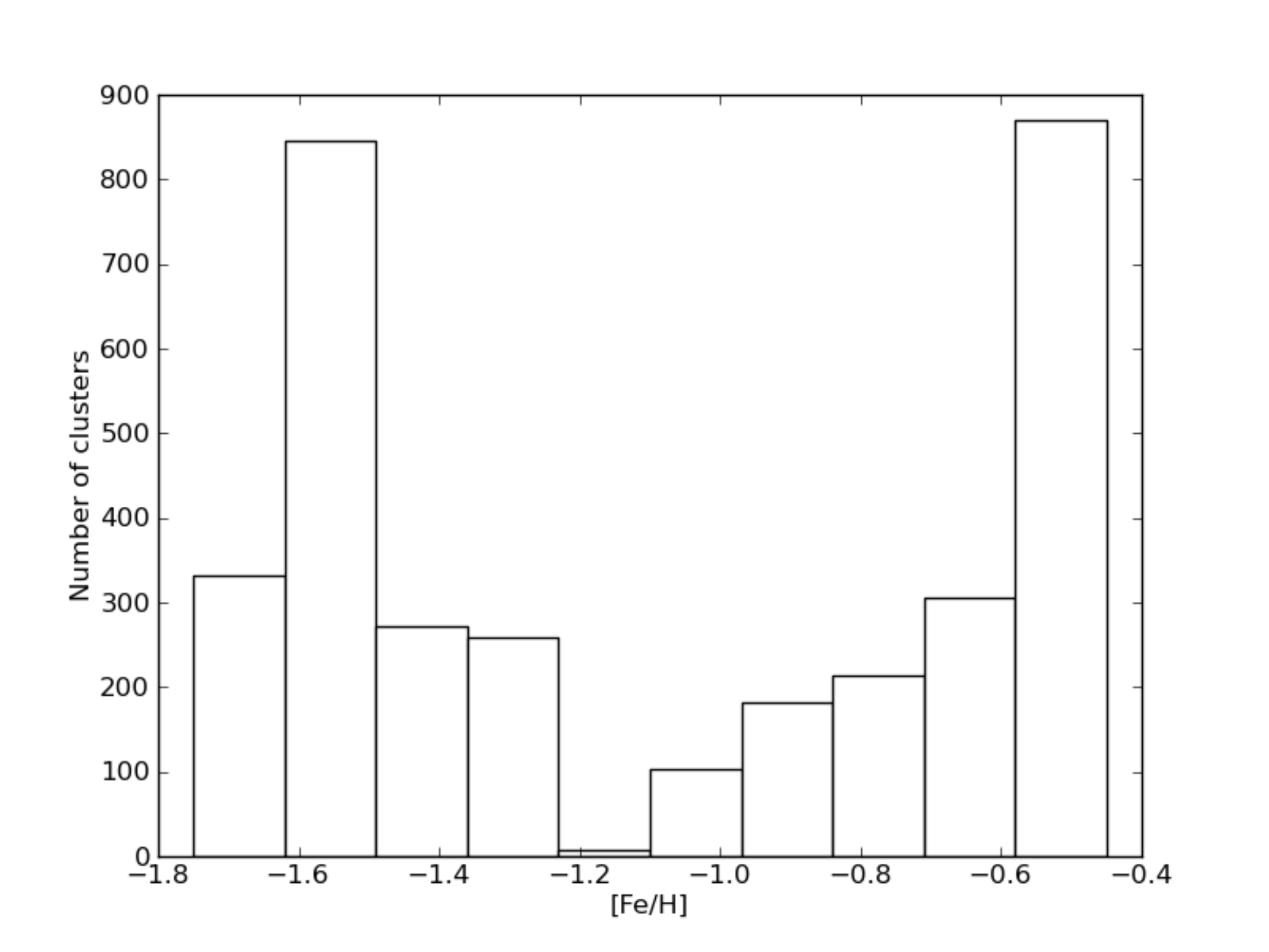}
  \caption{Metallicity histogram of Virgo globular clusters analyzed with GloMAGE using a catalog with cluster tidal radii of 160 pc.}
  \label{fig:virzdist160}
\end{figure}

Using GloMAGE, metallicities of these globular clusters have been estimated using g and z data published in the ACS Virgo Survey Catalog (\citet{Jordan09b}). The results are in part listed in Tab. \ref{tab:zcatvir}. As a cluster ID right ascension and declination are given conform with the ACS Survey in the J2000.0 system. The metallicity distribution is shown in Fig. \ref{fig:virzdist160}. It is obviously bimodal with peaks at $\mathrm{[Fe/H]}=-1.6$ and $\mathrm{[Fe/H]}=-0.5$, which agrees well with the results of \citet{Harris09}. Almost no clusters have metallicities between $\mathrm{[Fe/H]}=-1.1$ and $\mathrm{[Fe/H]}=-1.2$. That only half of the gaussian peak at $\mathrm{[Fe/H]}=-0.5$ is shown is caused by the high-metallicity boundary of the GloMAGE catalog.

\begin{figure}
  \centering
  \includegraphics[width=0.5\textwidth]{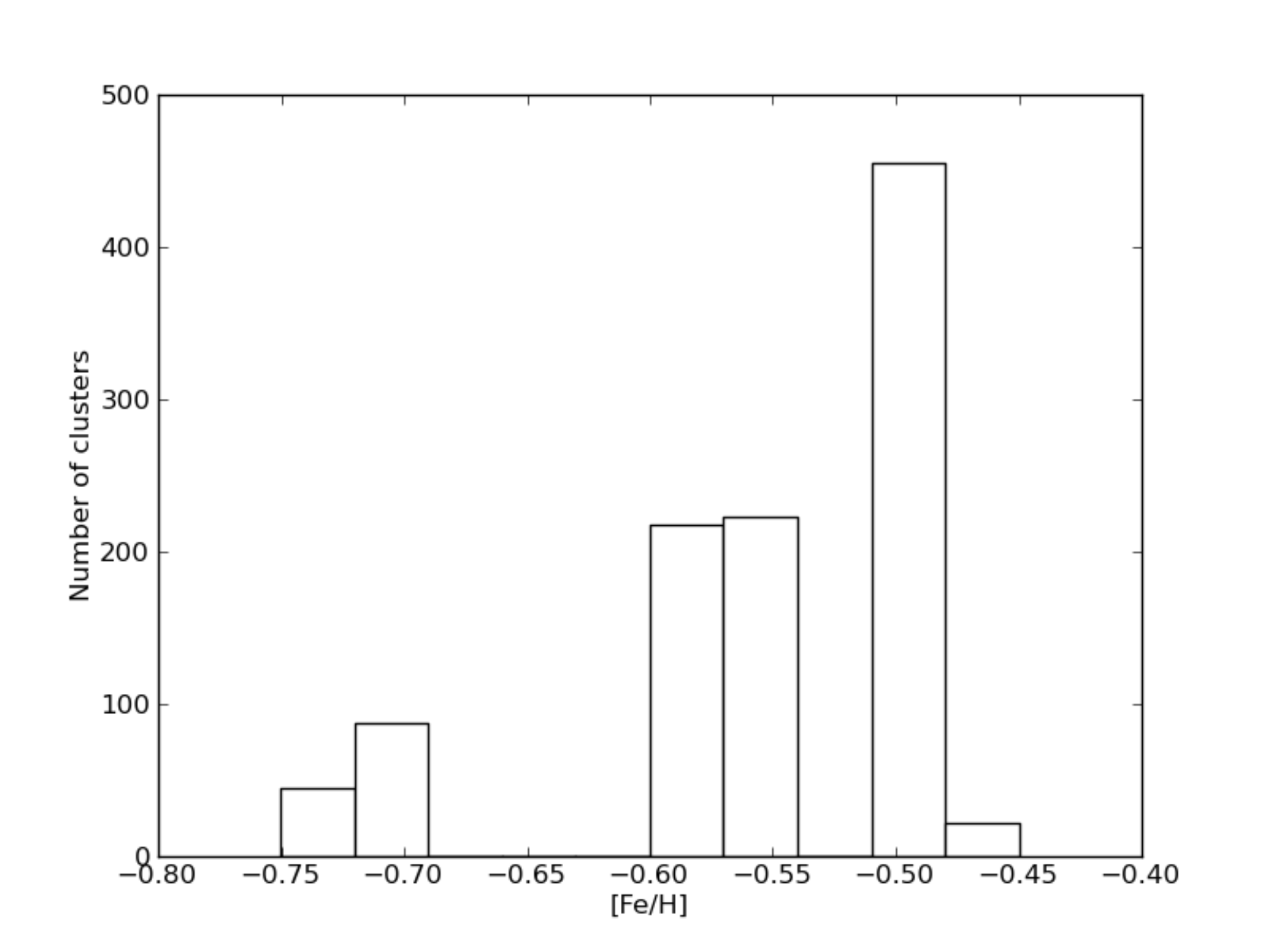}\\
  \includegraphics[width=0.5\textwidth]{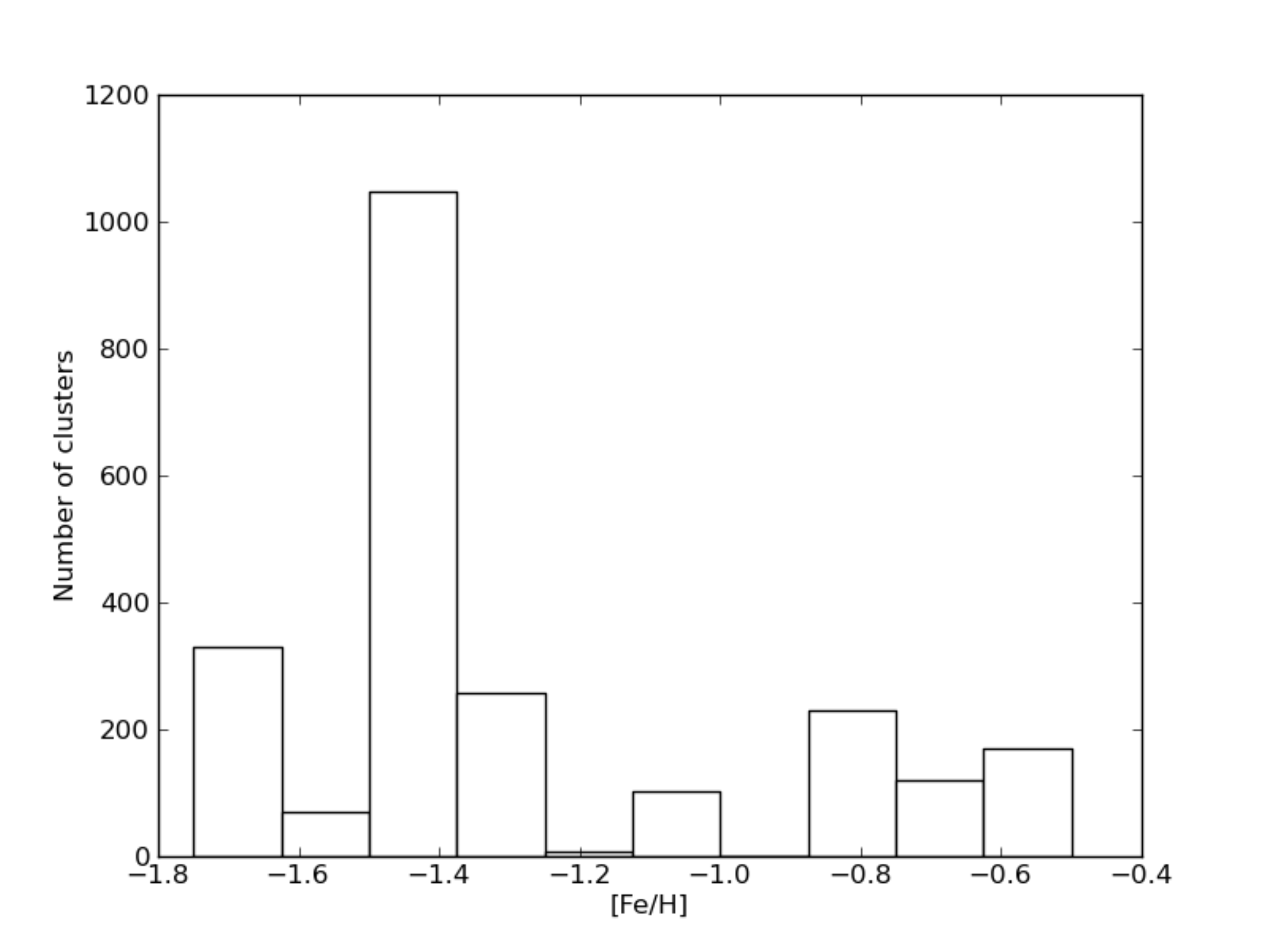}
  \caption{Metallicity distribution of Virgo Cluster globular clusters with colors $g-z>1.2$ (top) and $g-z<1.2$ (bottom).}
  \label{fig:virzdist-12}
\end{figure}

As seen in section \ref{sec:bimodality}, a metallicity distribution, not necessarily a bimodal distribution, leads to a bimodality in color. This system is an ideal laboratory to test whether the reverse is also the case, i.e. if a bimodal color distribution has to be caused by a bimodal metallicity distribution. Therefore, separately the clusters with colors on the blue and on the red side of $g-z=1.2$ have been analyzed and metallicity histograms have been produced. The result is shown in Fig. \ref{fig:virzdist-12}. In the case of $g-z>1.2$ (left figure) a peak at $\mathrm{[Fe/H]}=-0.5$ dominates the distribution with only a minor second peak at $\mathrm{[Fe/H]}=-0.7$, whereas in the case of $g-z<1.2$ there is a high peak at $\mathrm{[Fe/H]}=-1.4$ and a smaller second peak again at $\mathrm{[Fe/H]}=-0.7$. Therefore, the bimodal color distribution is caused by a bimodal metallicity distribution, the smaller second peak causing the minor third mode in Fig. \ref{fig:vircoldist}. Thus, the reverse conclusion, that a color bimodality is caused by a metallicity bimodality, is valid in this case. 

\begin{figure}
  \centering
  \includegraphics[width=0.5\textwidth]{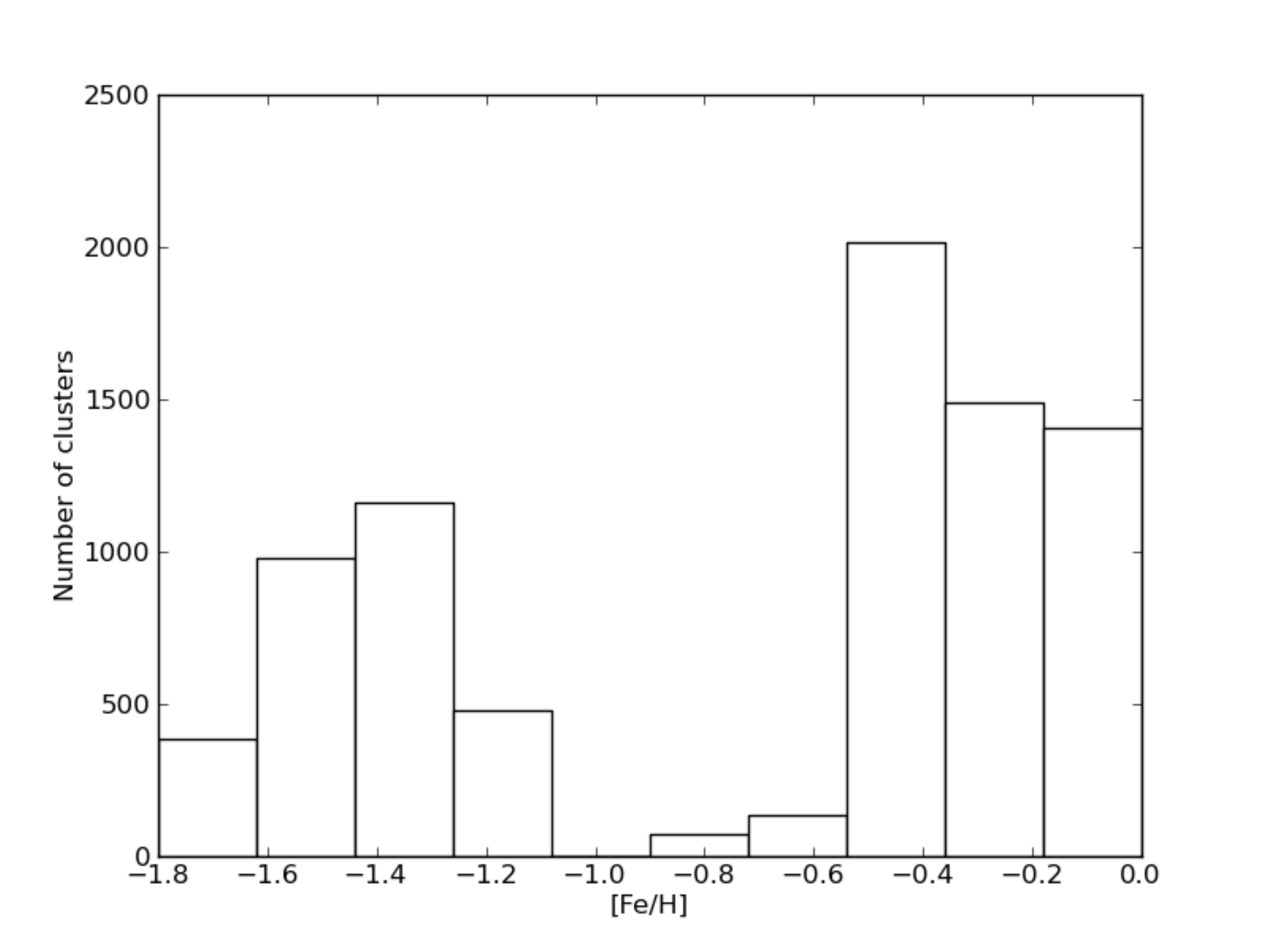}
  \caption{Metallicity histogram of Virgo globular clusters analyzed with GloMAGE using a catalog with cluster tidal radii of 80 pc.}
  \label{fig:virzdist80}
\end{figure}

Still, caution is needed, as the location of the peaks shifts with the influence of the tidal field. Fig. \ref{fig:virzdist80} shows the metallicity histogram of the Virgo globular clusters, but, instead of a tidal radius of 160 pc assumed for Fig. \ref{fig:virzdist160}, a tidal radius of only 80 pc is assumed. This causes the distribution to shift towards the red end of the histogram. A hypothetical spatial distribution of clusters covering a wide radial range around a galaxy could therefore be subject to very different tidal fields and cause a bimodality only through tidal interactions with the host galaxy. A second reason why the reverse conclusion of a color bimodality automatically implying a metallicity bimodality is wrong, is that by virtue of Fig. \ref{fig:zcol} a unimodal color distribution can be caused by a bimodal metallicity distribution. Therefore, a bimodal color distribution is allways caused by a bimodal metallicity distribution, but the reverse is not necessarily correct.

The dependence of the distinctiveness of the bimodality on cluster age explains the observation of \citet{Harris10a} in \ref{fig:harris10_02}, that low luminosity (and therefore old) cluster sets are clearly bimodal, whereas high luminosity (young) sets are not. Also the observations of \citet{Peng11} concerning the mean color values for different radial positions presented in the introduction can be interpreted, so that the inner clusters with a weaker bimodality are younger than the outer clusters. Also, the shifted peaks imply both different ages and different metallicity ranges and therefore different times of cluster formation.

The agreement of metallicity distribution functions of the Virgo and Coma galaxy clusters with GloMAGE shows that the color-metallicity relation derived in the previous chapter and tested here is applicable and agrees with previous metallicity determination.

\section{Summary}
\label{sec:conclusions}

Metallicity differences cause a color bimodality, as low-metallicity clusters are significantly bluer than high-metallicity clusters, but the transformation of metallicity to color is nonlinear for metallicities below $\log{z}\approx -2.2$. This nonlinearity is caused by the appearance of Red Horizontal Branch stars and an elongation of the Asymptotic Giant Branch.

This information was used to compile the GloMAGE catalog of 100 metallicities between solar and 1\% solar metallicity, a tidal radius of 160 pc, a binary fraction of 10\%, a King profile parameter of $W_0=6$ and a time step of 50 Myr. The package reads ugriz colors for any observation and returns the metallicity of the observed cluster. Results for M31 agree well with observations and data of the ACS Virgo Cluster Survey was used to explain the bimodal color distribution of the globular cluster population in the Virgo cluster.

Several predictions can be made:
\begin{enumerate}
  \item The binary fraction and the density profile of a cluster are not important for their integrated properties.
  \item The bimodality is an effect of a nonlinearity in the color-metallicity relation.
  \item The \lq life-time\rq\, of the bimodality depends on metallicity and therefore on stellar evolution processes.
  \item Colors including only the UV show a weaker bimodality than those subtracting from visual bands.
  \item Cluster sets with a distinct bimodality are in principle older than those with only a weak bimodal distribution.
  \item A bimodal color distribution of coeval clusters implies a bimodal metallicity distribution, but a unimodal color distribution does not imply a unimodal metallicity distribution.
  \item A non-coeval cluster set can show a color bimodality with a unimodal metallicity distribution.
  \item The tidal field can shift the peaks of histogram modes and therefore cause a bimodal color distribution.
\end{enumerate}

Future work will include applying this scheme to a measure for age, for instance the mass-to-light ratio.

\section*{Acknowledgements}
I thank Rainer Spurzem and Jonathan M. B. Downing for scientific advice and supervision. I acknowledge the National Astronomical Observatories of China and the Chinese Academy of Sciences for funding.

\bibliography{paper}{}
\bibliographystyle{mn2e}

\label{lastpage}

\end{document}

%% file: cat_m31_part.txt
B054	&	-2	&	0.9	&	-1.9	&	0.1	\\
B154	&	-2.05	&	0.2	&	-1.9	&	0.2	\\
B029	&	-2.3	&	0.4	&	-2	&	0.1	\\
B150	&	-2.35	&	0.4	&	-2	&	0.1	\\
B153	&	-2	&	0.7	&	-2	&	0.1	\\
B164	&	-2.3	&	0.5	&	-2	&	0.2	\\
B171	&	-2.3	&	0.5	&	-2	&	0.1	\\
B197	&	-2.5	&	0.5	&	-2	&	0.1	\\
B379	&	-2.2	&	0.3	&	-2.1	&	0.1	\\
B398	&	-2	&	1	&	-2.1	&	0.2	\\

%% file: cat_vir_part.txt
187.4469413\_7.9997839  &  -2.15  &  0.3\\
187.4454712\_8.0026666  &  -3.0  &  0.2\\
187.4474911\_8.0010406  &  -2.55  &  0.3\\
187.4423850\_7.9987706  &  -3.2  &  0.3\\
187.4471733\_8.0027498  &  -2.15  &  0.3\\
187.4479377\_7.9985460  &  -2.2  &  0.3\\
187.4459688\_8.0050825  &  -3.1  &  0.1\\
187.4407561\_8.0032889  &  -3.2  &  0.1\\
187.4495998\_8.0014389  &  -3.2  &  0.1\\
187.4517469\_7.9988995  &  -2.25  &  0.1\\